\documentclass[structabstract]{aa}
\usepackage[utf8]{inputenc}
\usepackage{txfonts}
\usepackage{graphicx}
\usepackage{xcolor}
\usepackage{natbib}
\usepackage{multirow}
\usepackage{upgreek}

\bibpunct{(}{)}{;}{a}{}{,}

\newcommand{\ie}{{\em i.e.,}}
\newcommand{\eg}{{\em e.g.,}}

\newcommand{\emm}[1]{\ensuremath{#1}}
\newcommand{\emr}[1]{\emm{\mathrm{#1}}}

\newcommand{\unit}[1]{\emr{\,#1}}

\newcommand{\Av}{\emm{A_\emr{V}}}
\newcommand{\Ak}{\emm{A_\emr{K}}}

\newcommand{\kHz}{\unit{kHz}}

\newcommand{\GHz}{\unit{GHz}}
\newcommand{\Myr}{\unit{Myr}}
\newcommand{\Msun}{\unit{M_{\sun}}}
\newcommand{\pc}{\unit{pc}}
\newcommand{\mpc}{\unit{mpc}}
\newcommand{\pccm}{\unit{cm^{-3}}} % per cubic centimeter
\newcommand{\magn}{\unit{mag}}
\newcommand{\micron}{\unit{\upmu m}}

\newcommand{\K}{\unit{K}}
\newcommand{\kms}{\unit{km\,s^{-1}}}

\newcommand{\Hii}{\textsc{Hii}}

\newcommand{\FigCloud}{%
  \begin{figure}
    \centering
    \includegraphics[width=\linewidth]{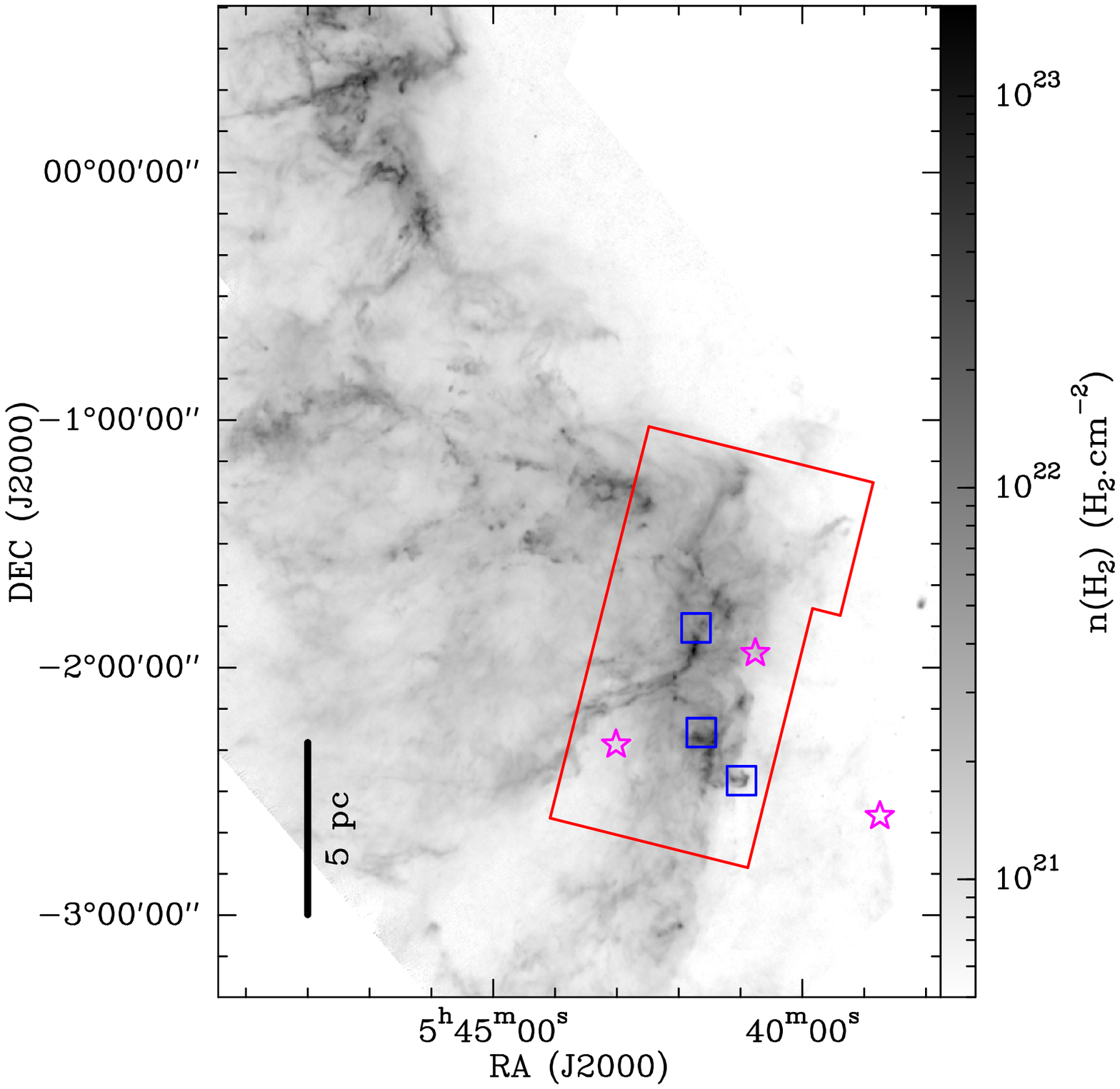}
    \caption{H$_2$ column density map of the south-western part of the Orion\,B giant molecular cloud, derived from \emph{Herschel Gould Belt Survey} observations \citep{andre10,schneider13}. The field observed by the Orion\,B collaboration and used for this work is overlaid in red. The blue squares mark the nebulae NGC\,2024, NGC\,2023 and the Horsehead, and the pink star symbols mark the stars Alnitak, HD\,38087 and $\sigma$\,Ori (North to South).}
    \label{fig:cloud}
  \end{figure}}

\newcommand{\FigChannelMaps}{%
  \begin{figure*}
    \centering
    \includegraphics[height=0.25\paperheight]{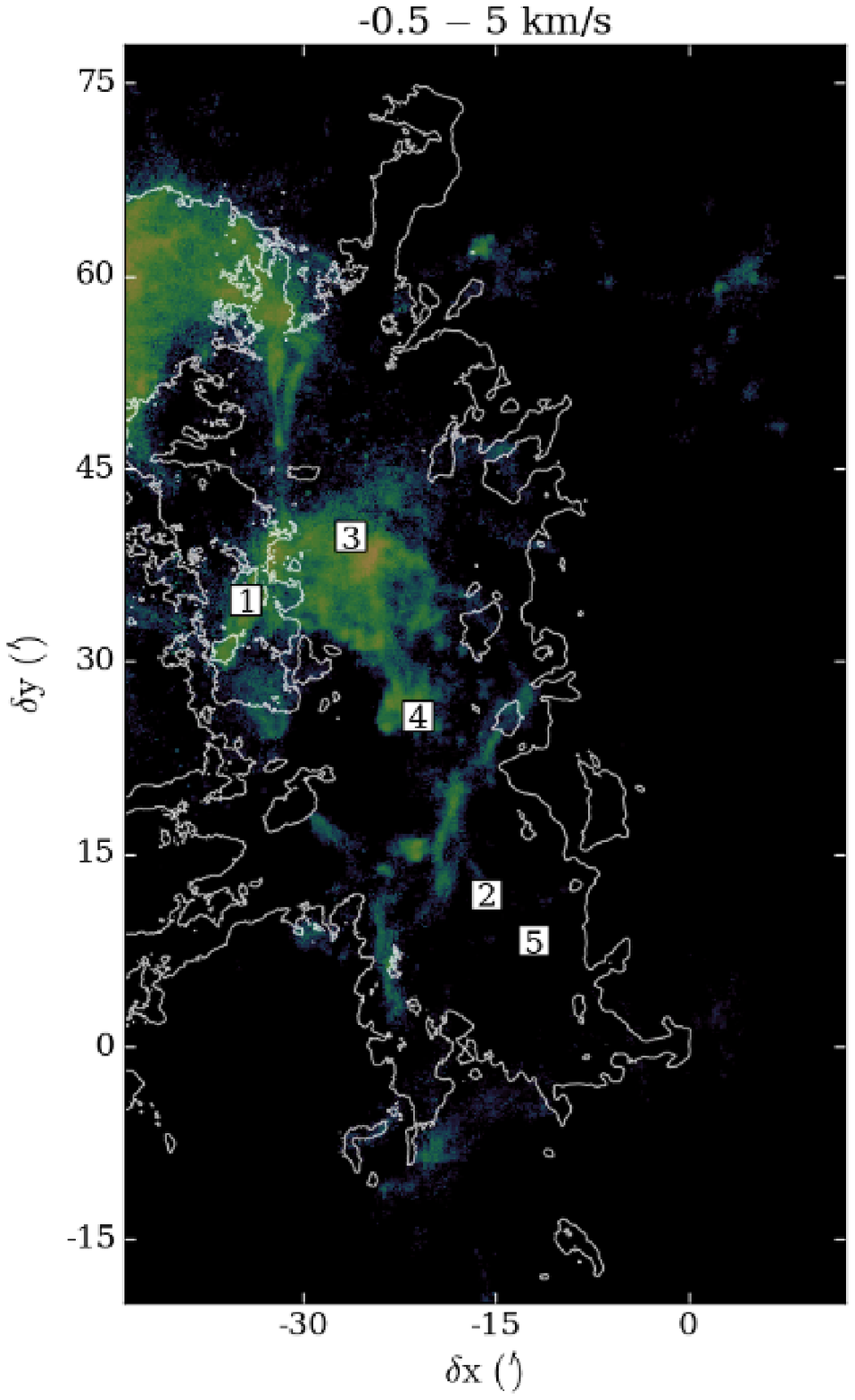}
    \includegraphics[height=0.25\paperheight]{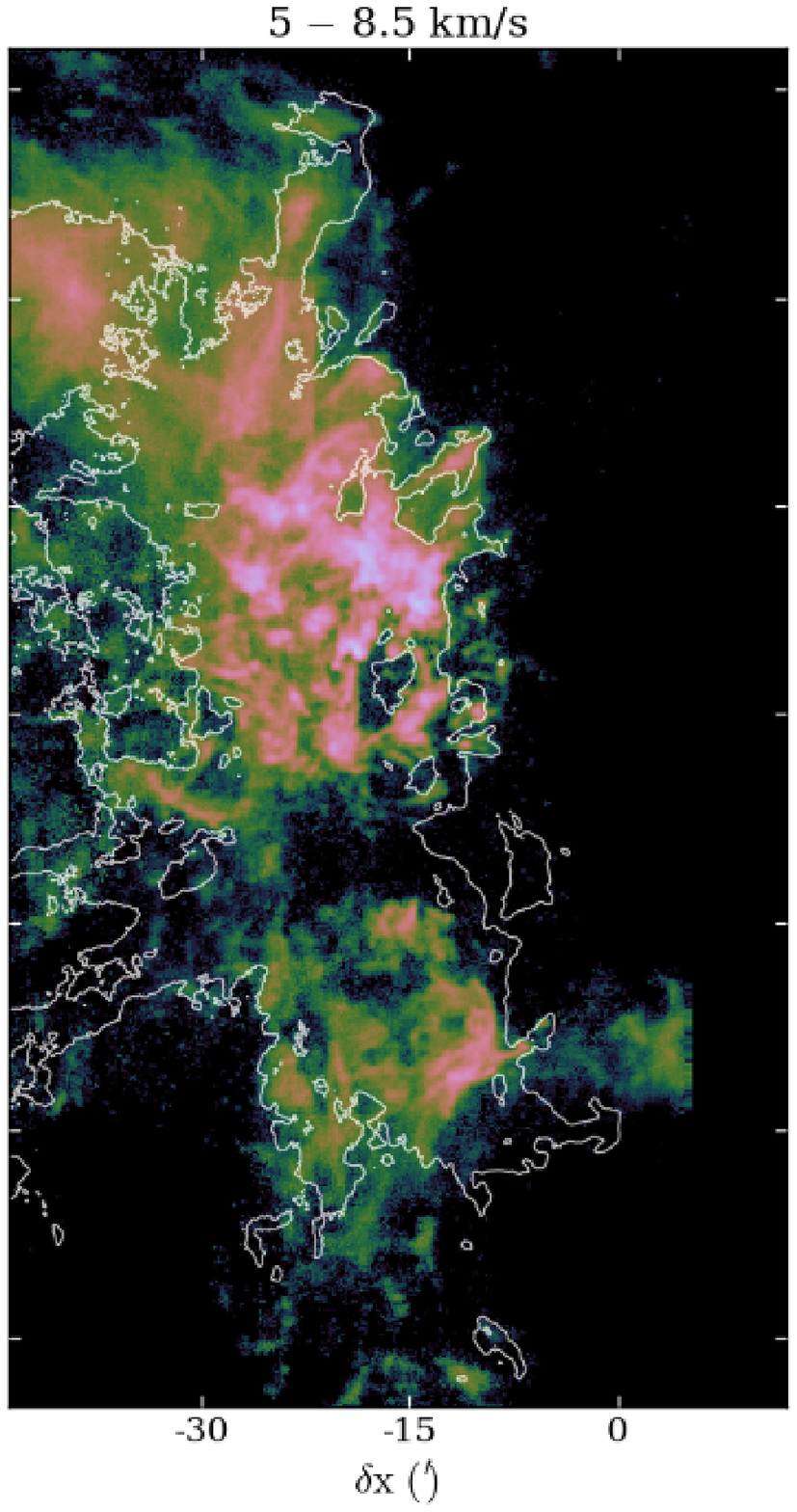}
    \includegraphics[height=0.25\paperheight]{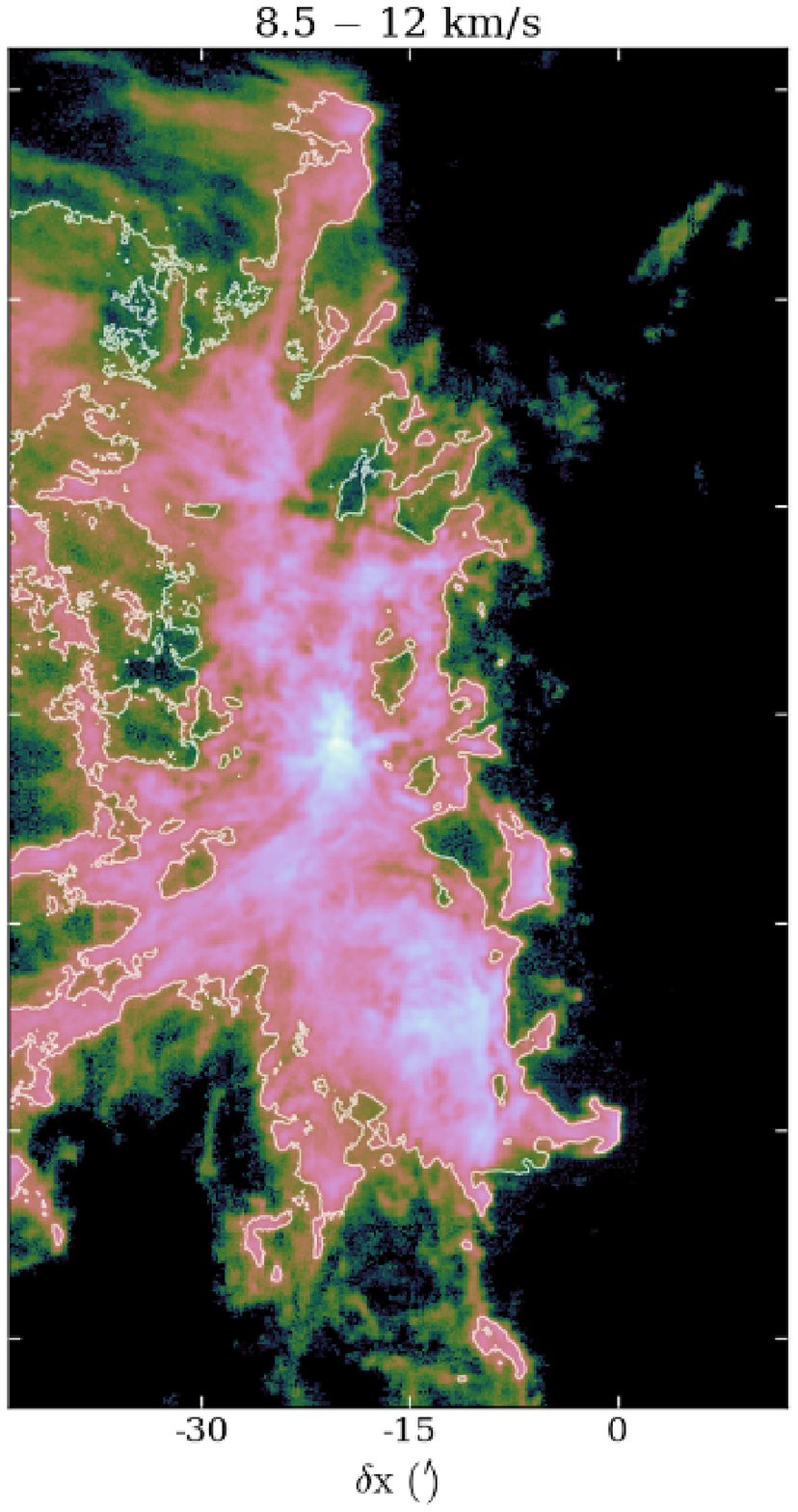}
    \includegraphics[height=0.25\paperheight]{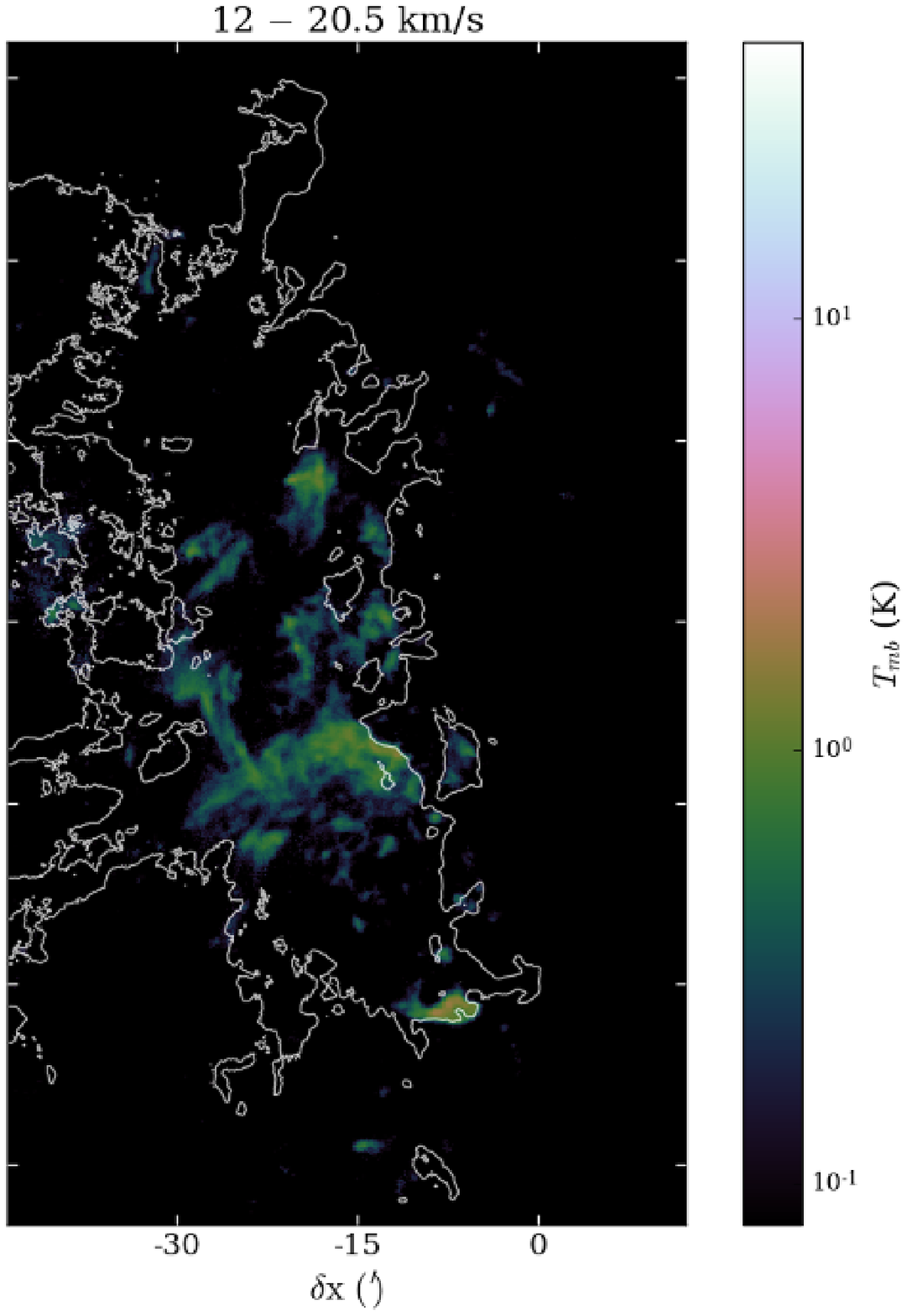}
    \caption{Maps of the average brightness temperature of the  $^{13}$CO($J=1-0$) line in four contiguous velocity ranges. The main-beam temperature scale is indicated by the color bar on the right. The contour shows the value of 8.9\,K\,\kms{} in the $W_0$ map, corresponding to 0.43\,K in the mean temperature map integrated over the -0.5 -- 20.5\kms{} velocity range. The set of coordinates used for the observational campaign takes the Horsehead PDR as a reference point, and aligns the IC\,434 PDR along the vertical axis (14$^\circ$ counter-clockwise rotation with respect to equatorial coordinates). The numbered squares in the first panel show the positions of the spectra presented in Fig.~\ref{fig:spectra}, from left to right.}
    \label{fig:channelmaps}
  \end{figure*}}

\newcommand{\FigSpectra}{%
  \begin{figure*}
    \centering
    \includegraphics[width=\linewidth]{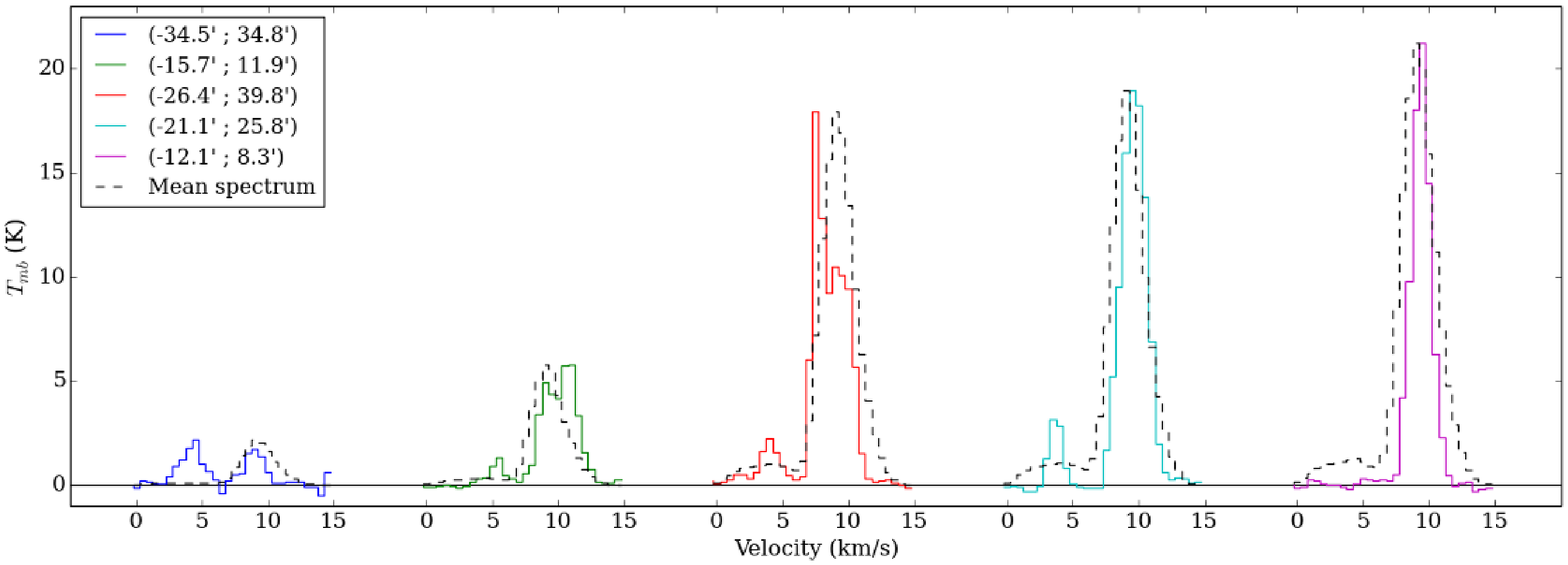}
    \caption{$^{13}$CO$(J=1-0)$ spectra along selected lines of sights in the Orion\,B cloud, showing the diversity of velocity components (up to four per spectrum). The coordinates of the various lines of sights are given in arc-minutes in our custom set of coordinates ($\delta$x, $\delta$y). The average spectrum for the whole field, normalized to have the same peak temperature, is superimposed with a dashed line for comparison. The position of the spectra on the map are shown by white squares in the first panel of Fig.~\ref{fig:channelmaps}.}
    \label{fig:spectra}
  \end{figure*}}

\newcommand{\FigWzero}{%
  \begin{figure*}
    \centering
    \includegraphics[width=0.49\linewidth]{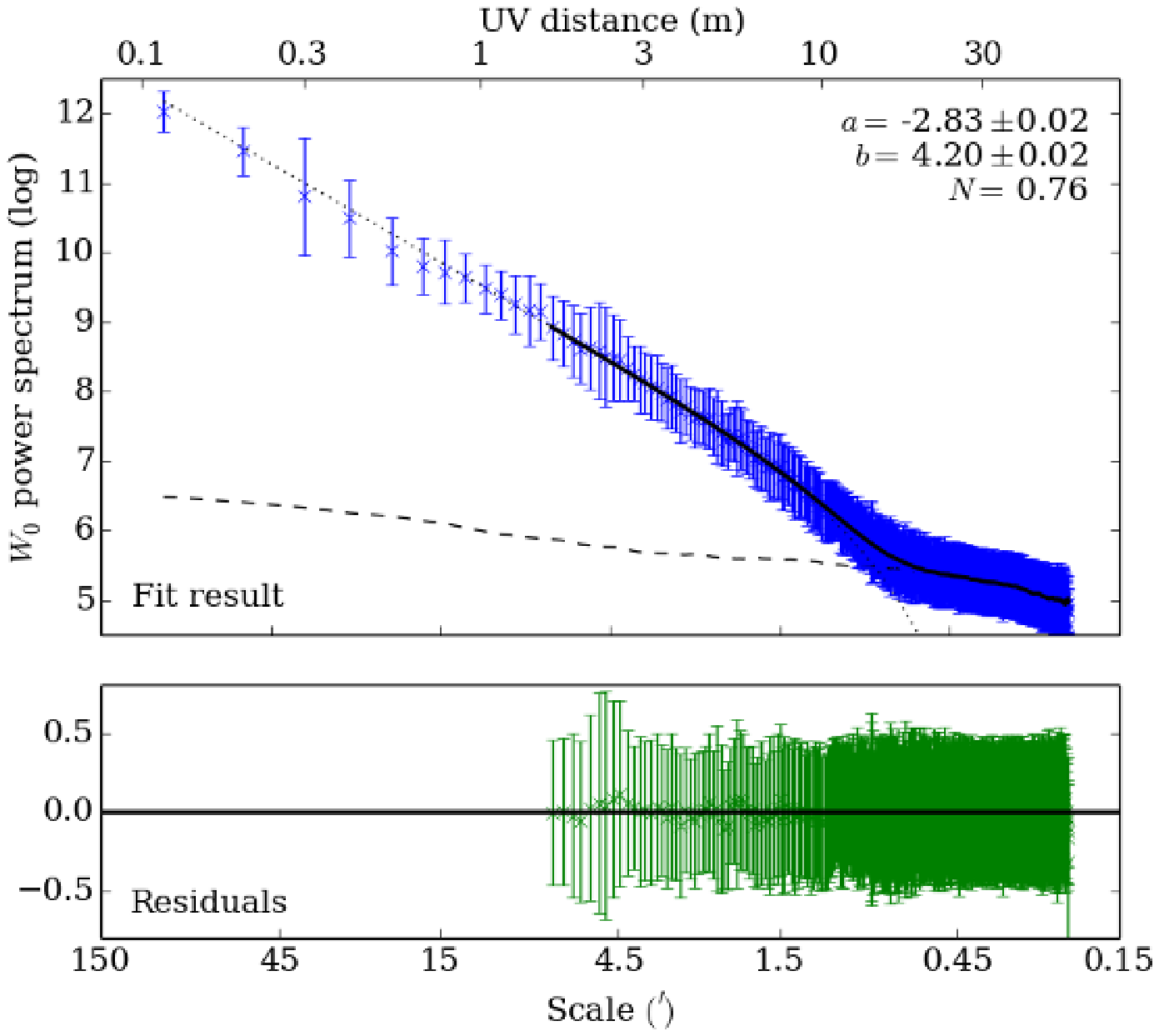}
    \includegraphics[width=0.49\linewidth]{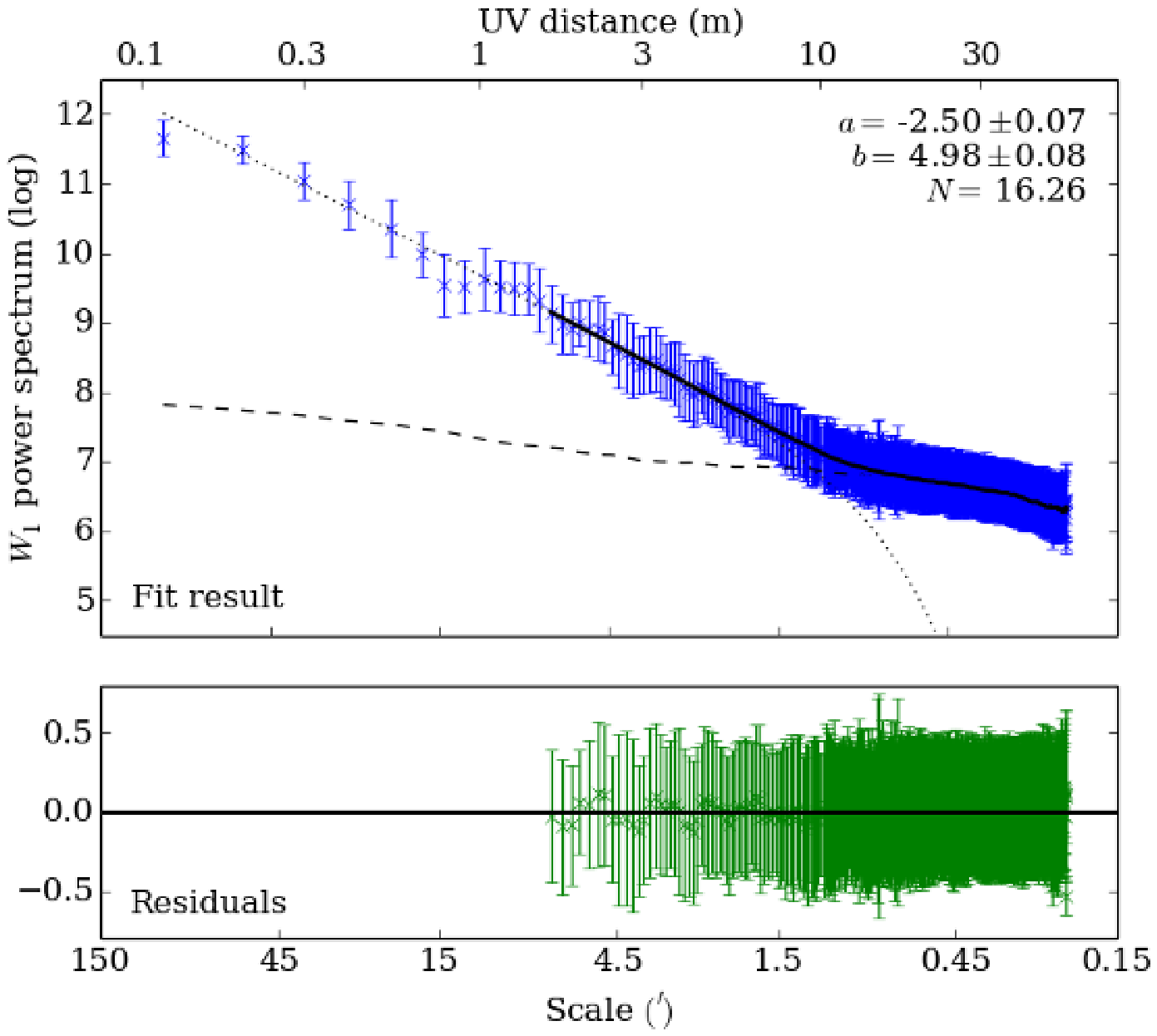}
    \caption{Left: Power law fitting of the $W_0$ power spectrum. Upper panel: data (blue crosses), fit result (thick solid line) plotted over the fitted domain, power law convolved with the Gaussian beam (dotted line) extrapolated to all spatial frequencies, noise model (dashed line). Lower panel, residuals. The scale in the Fourier space is given in UV distance as for interferometric observations, which allows to visually relate the resolution and the telescope diameter. Right: Same results, except for the $W_1$ power spectrum.}
    \label{fig:W0}
  \end{figure*}}

\newcommand{\FigMachImage}{%
  \begin{figure}
    \centering
    \includegraphics[height=0.22 \paperheight]{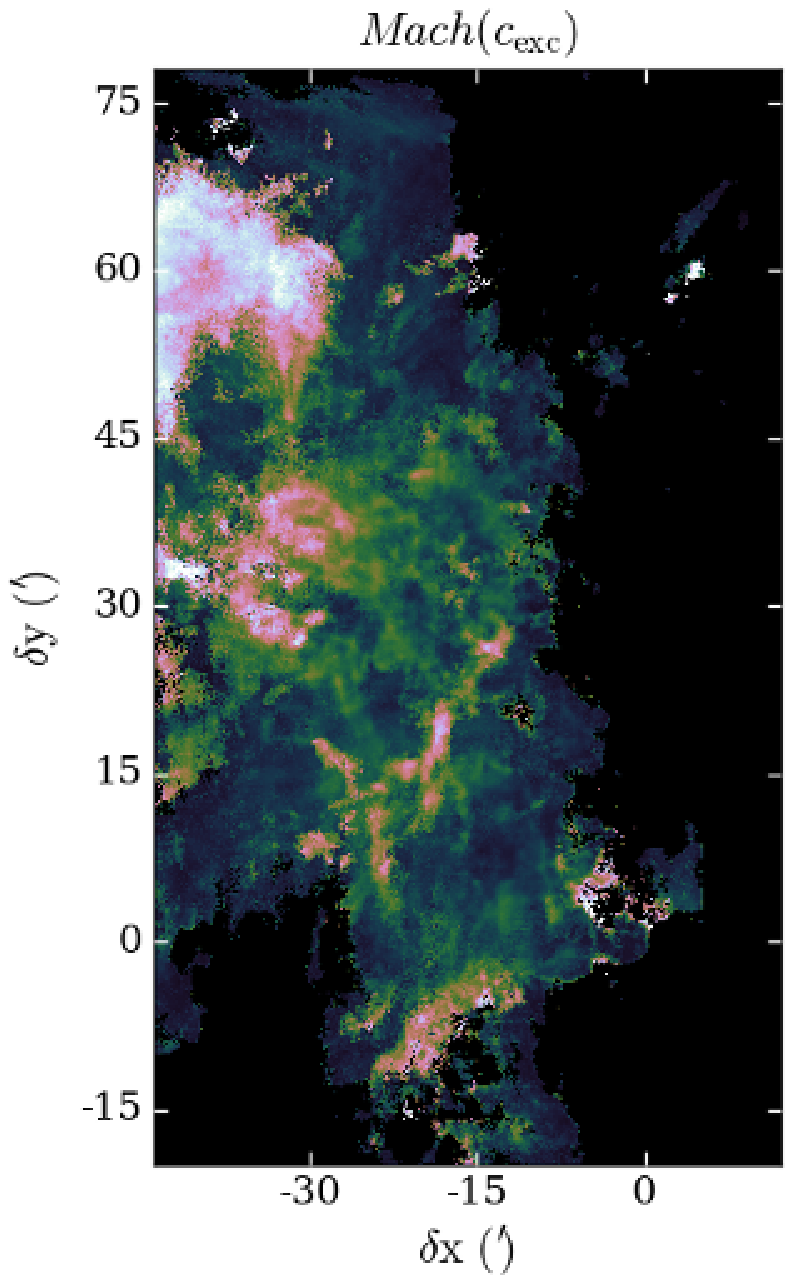}
    \includegraphics[height=0.22 \paperheight]{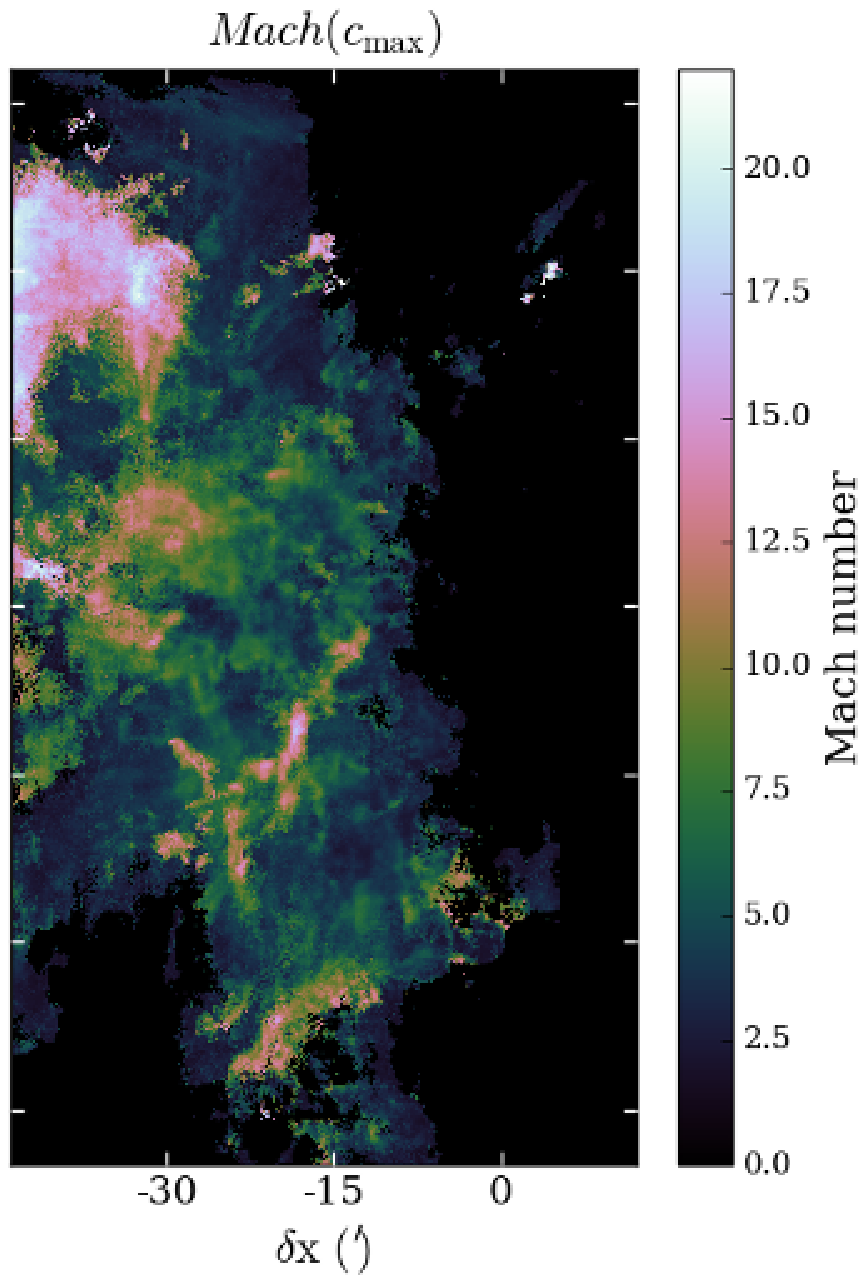}
    \caption{Mach number maps computed using the two different temperature maps, left using $T_\emr{exc}$, right using $T_\emr{max}$}
    \label{fig:mach:ima}
  \end{figure}}

\newcommand{\FigMachHist}{%
  \begin{figure}
    \centering
    \includegraphics[width=\linewidth]{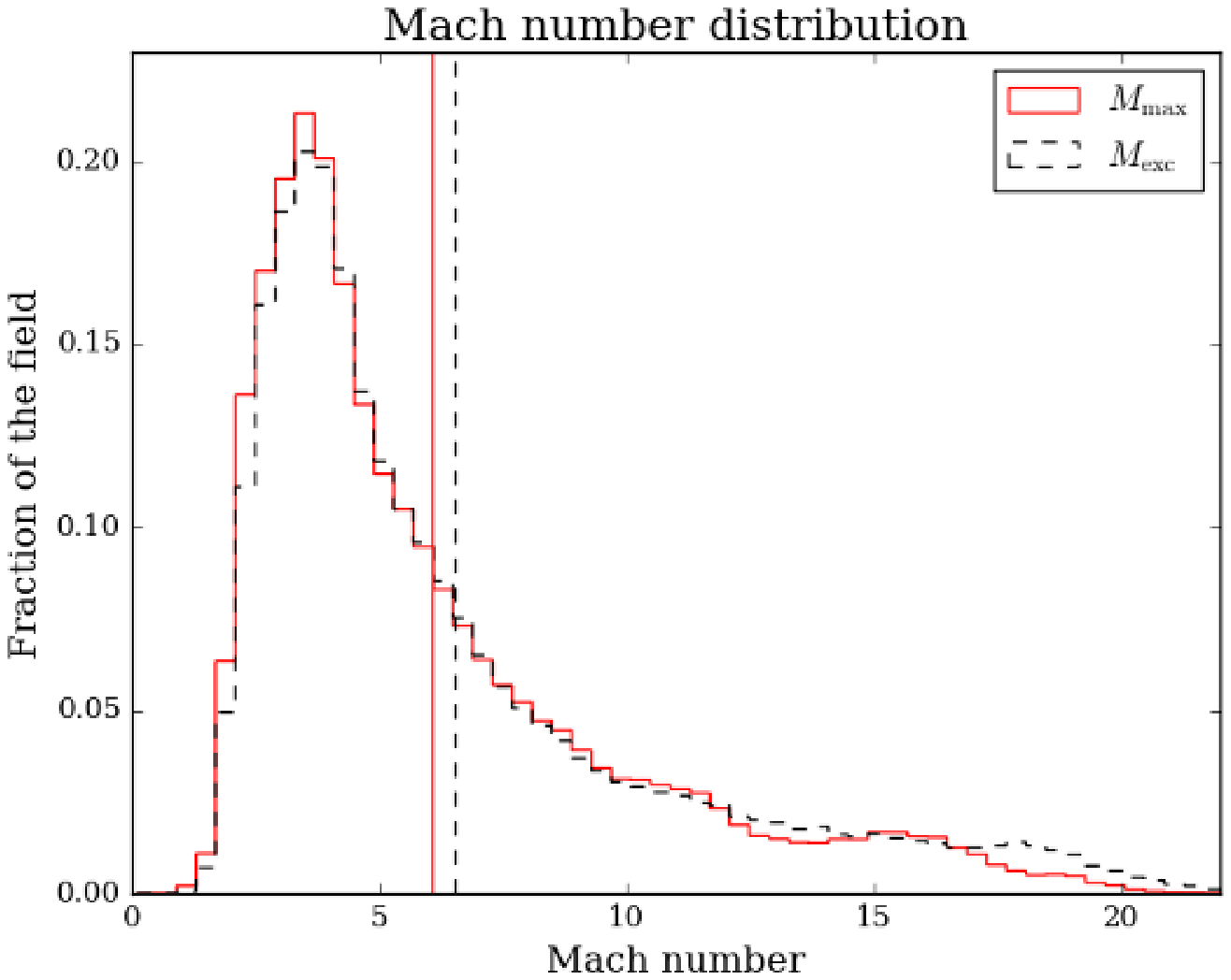}
    \caption{Histogram of the Mach number map, for $T_\emr{max}$ ant $T_\emr{exc}$. The vertical bars show the position of the mean value of each distribution.}
    \label{fig:mach:hist}
  \end{figure}}

\newcommand{\FigSoundSpeed}{%
  \begin{figure*}
    \centering
    \includegraphics[height=0.25\paperheight]{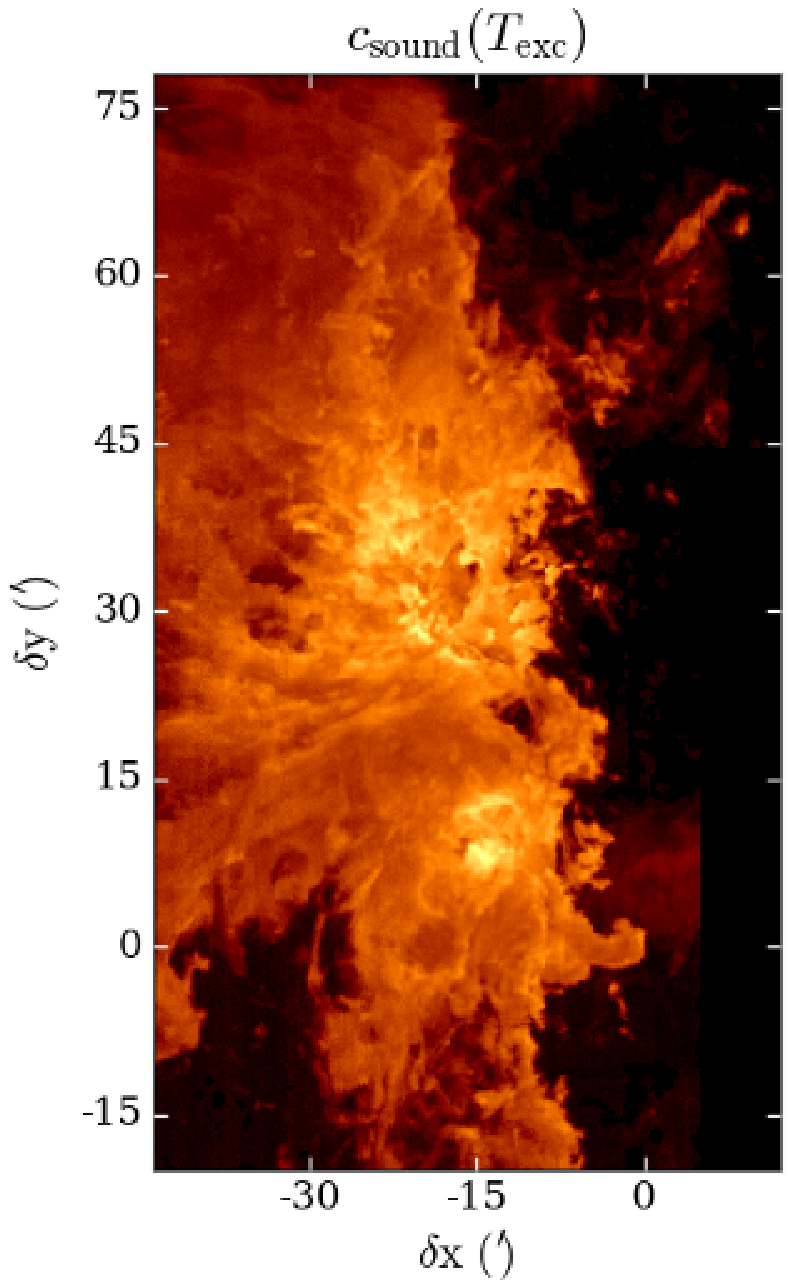}
    \includegraphics[height=0.25\paperheight]{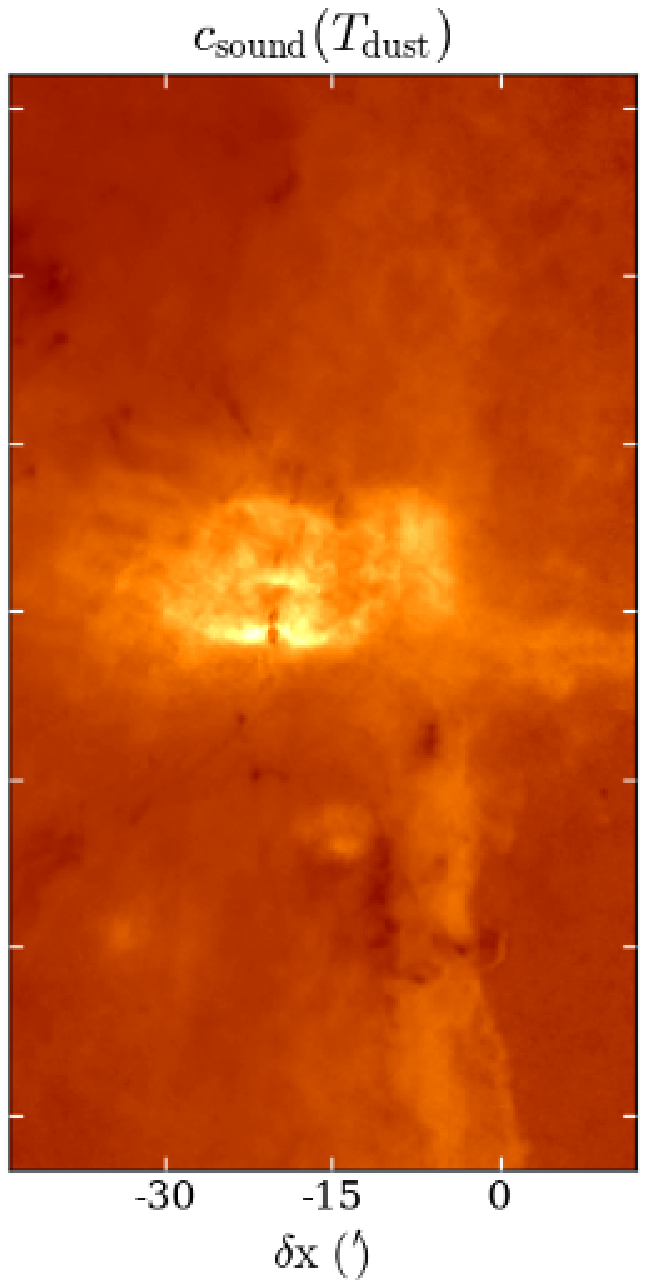}
    \includegraphics[height=0.25\paperheight]{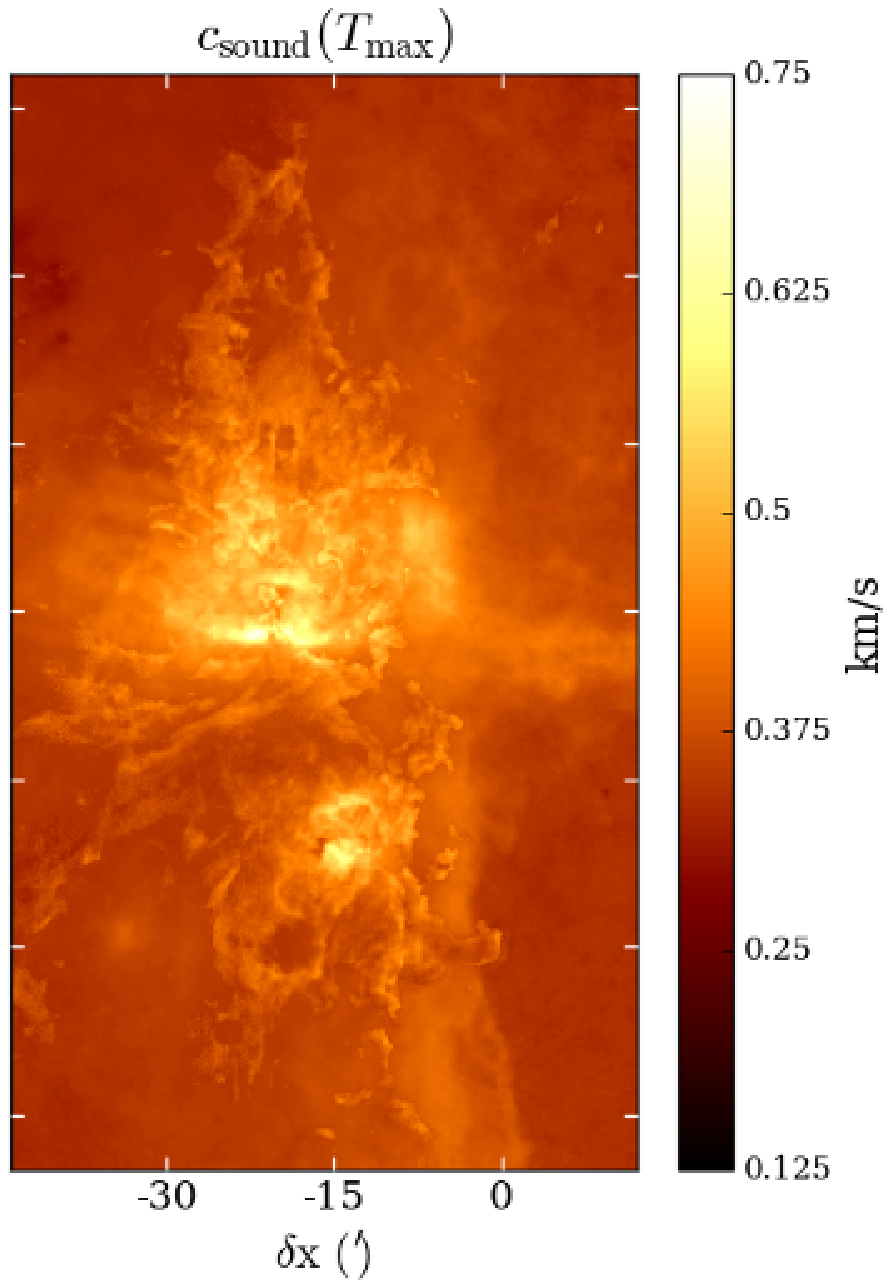}
    \includegraphics[height=0.25\paperheight]{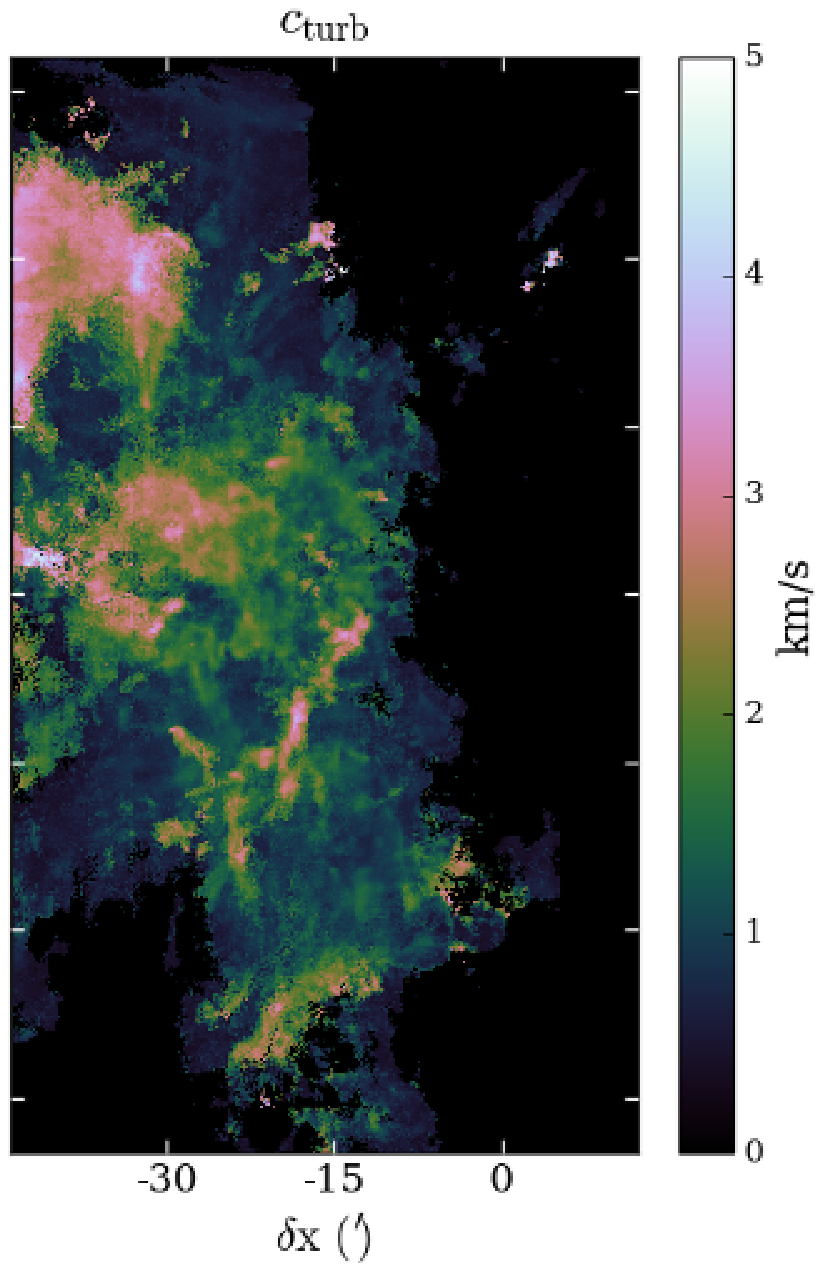}
    \caption{First three panels: speed of sound computed for the estimated excitation temperature of the gas, the dust temperature, and the maximum of the two previous ones. Last panel: turbulent flow velocity dispersion. We assume that having one or several spectral components along the line of sight is irrelevant, since one component gives a turbulent velocity dispersion, and two components show a high-velocity shock.}
    \label{fig:soundspeed}
  \end{figure*}}

\newcommand{\FigZoom}{%
  \begin{figure*}
    \centering
    \includegraphics[height=0.2\paperheight]{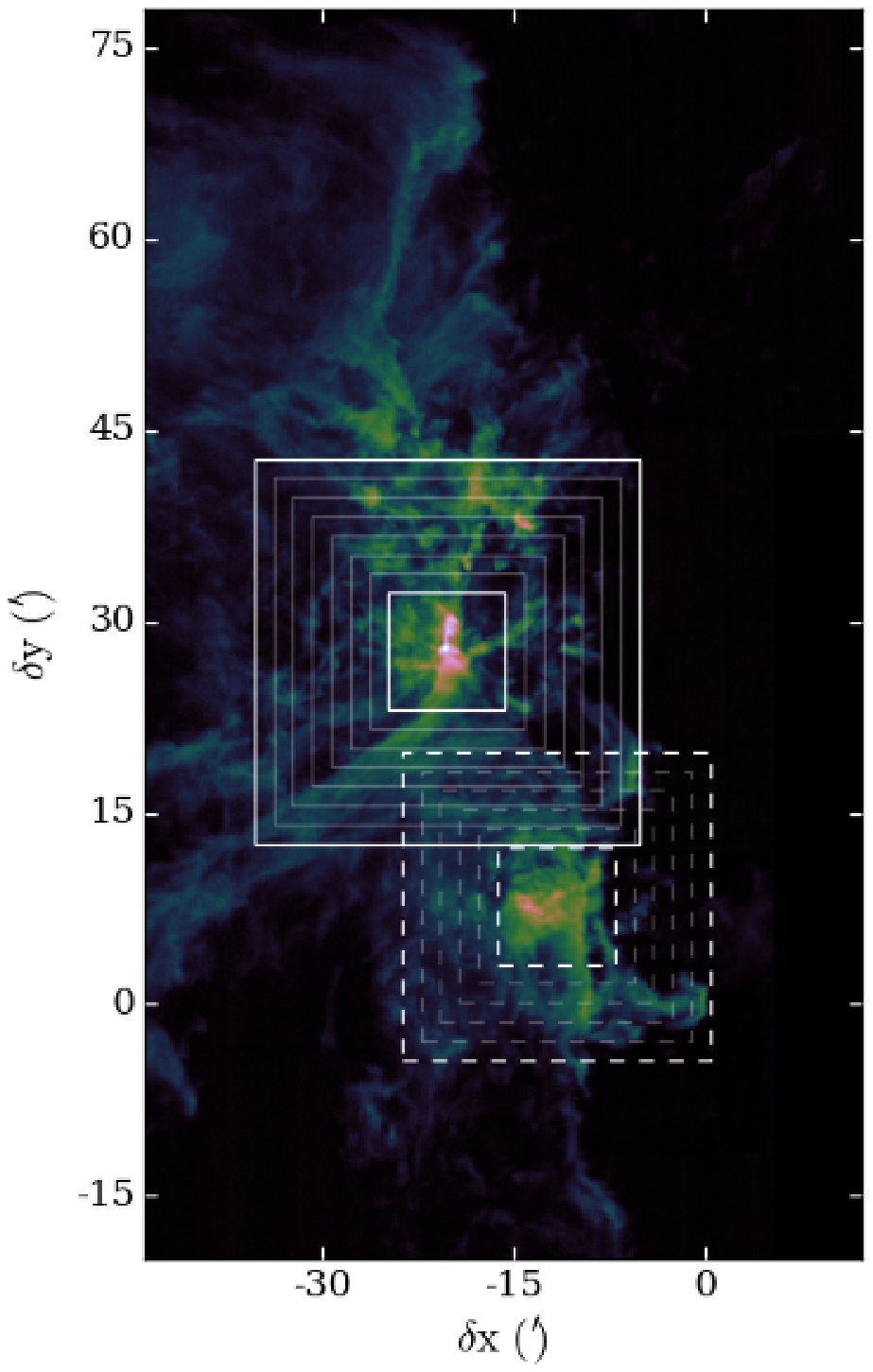}
    \includegraphics[height=0.2\paperheight]{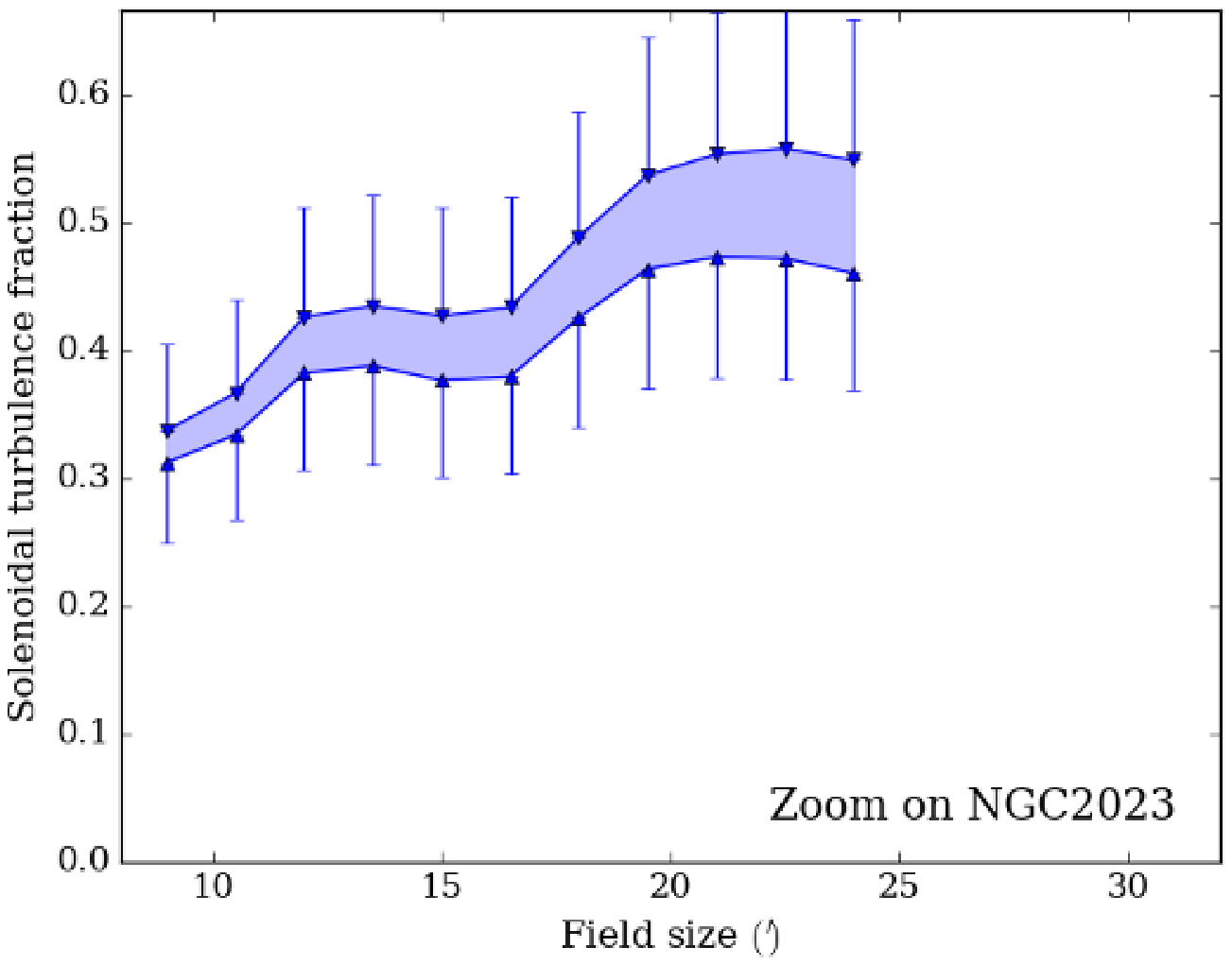}
    \includegraphics[height=0.2\paperheight]{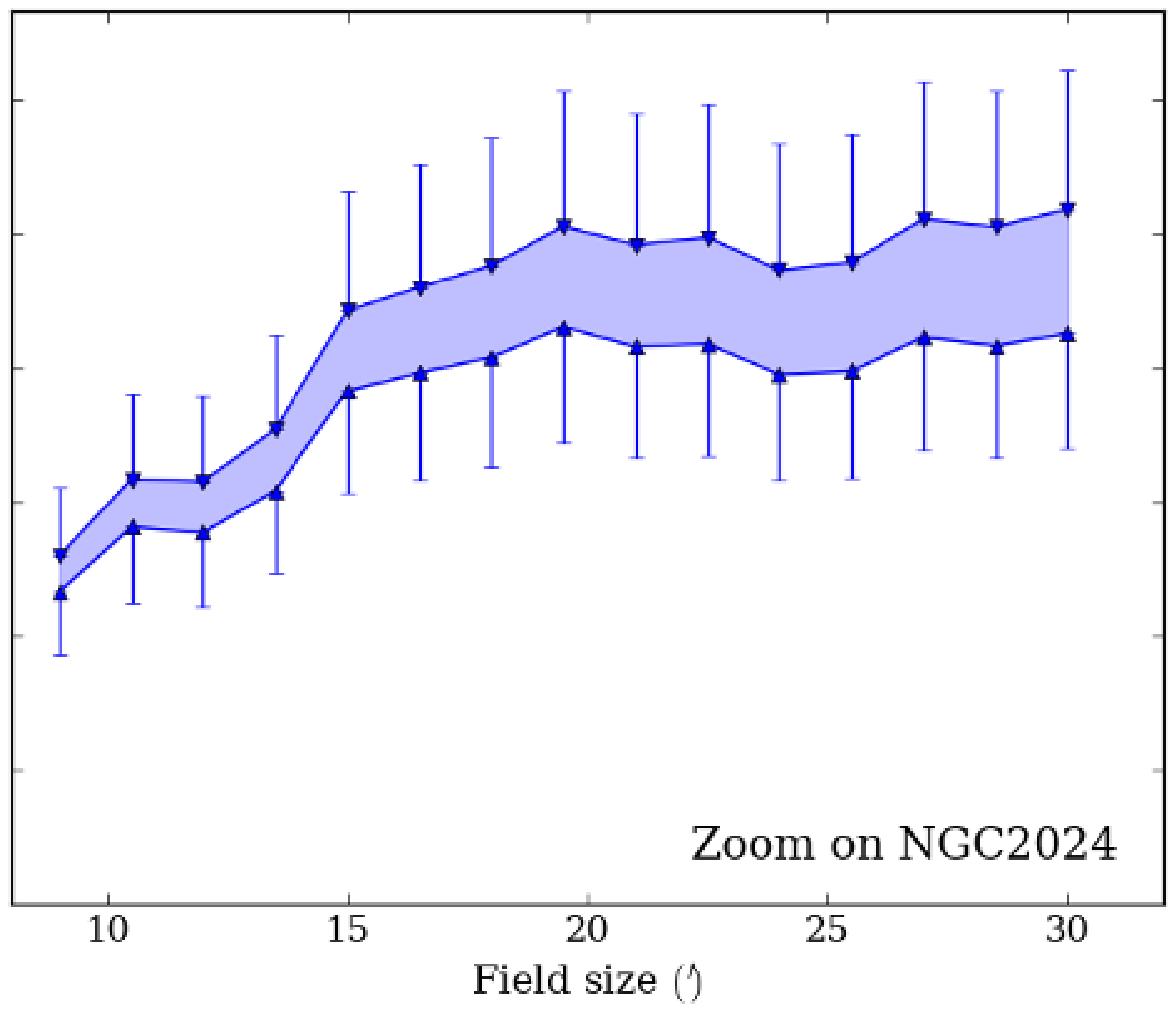}
    \caption{Solenoidal turbulence fraction for zooms on the NGC\,2023 and NGC\,2024 star-forming regions. The shaded area corresponds to the $g_{21}$ uncertainty, while the error bars show the experimental uncertainties due to observational noise. The upper limit of the plots marks the equipartition limit. The map on the left presents the areas for which the calculations were performed (solid squares: NGC\,2024, dashed squares: NGC\,2023), superimposed on the $^{13}$CO($J=1-0$) $T_\emr{peak}$ map.}
    \label{fig:zoom}
  \end{figure*}}
    
\newcommand{\FigSlide}{%
  \begin{figure}
    \centering
    \includegraphics[width=\linewidth]{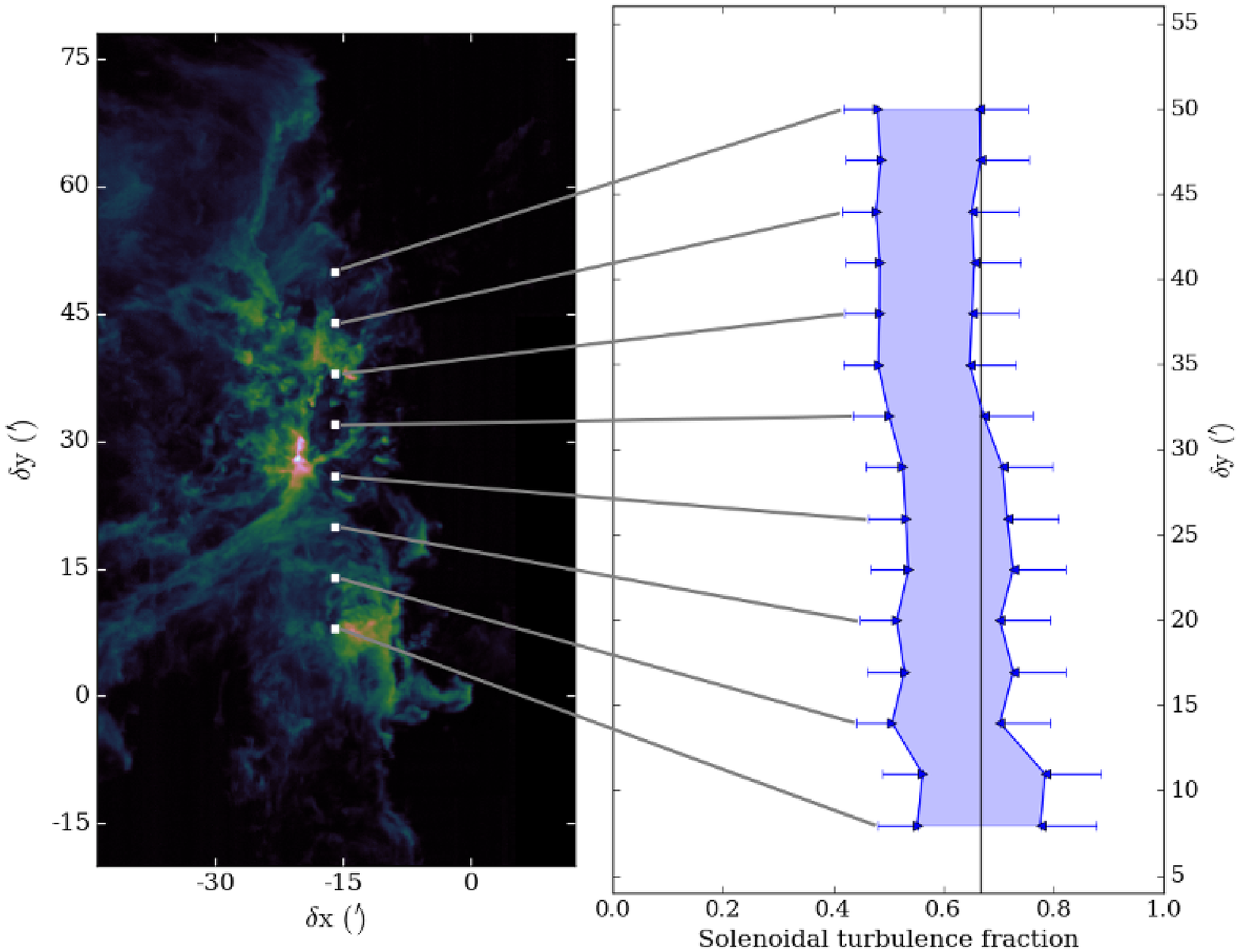}
    \caption{Solenoidal turbulence fraction for sliding square areas with a width equal to the one of the full map (56 arcminutes). The shaded area corresponds to the $g_{21}$ uncertainty, while the error bars show the experimental uncertainties due to observational noise. The vertical line marks the equipartition limit. The map on the left presents the centres of the square areas for which the calculations were performed, superimposed on the $^{13}$CO($J=1-0$) $T_\emr{peak}$ map.}
    \label{fig:slide}
  \end{figure}}  

\newcommand{\FigMoments}{%
  \begin{figure*}
    \centering
    \includegraphics[width=\linewidth]{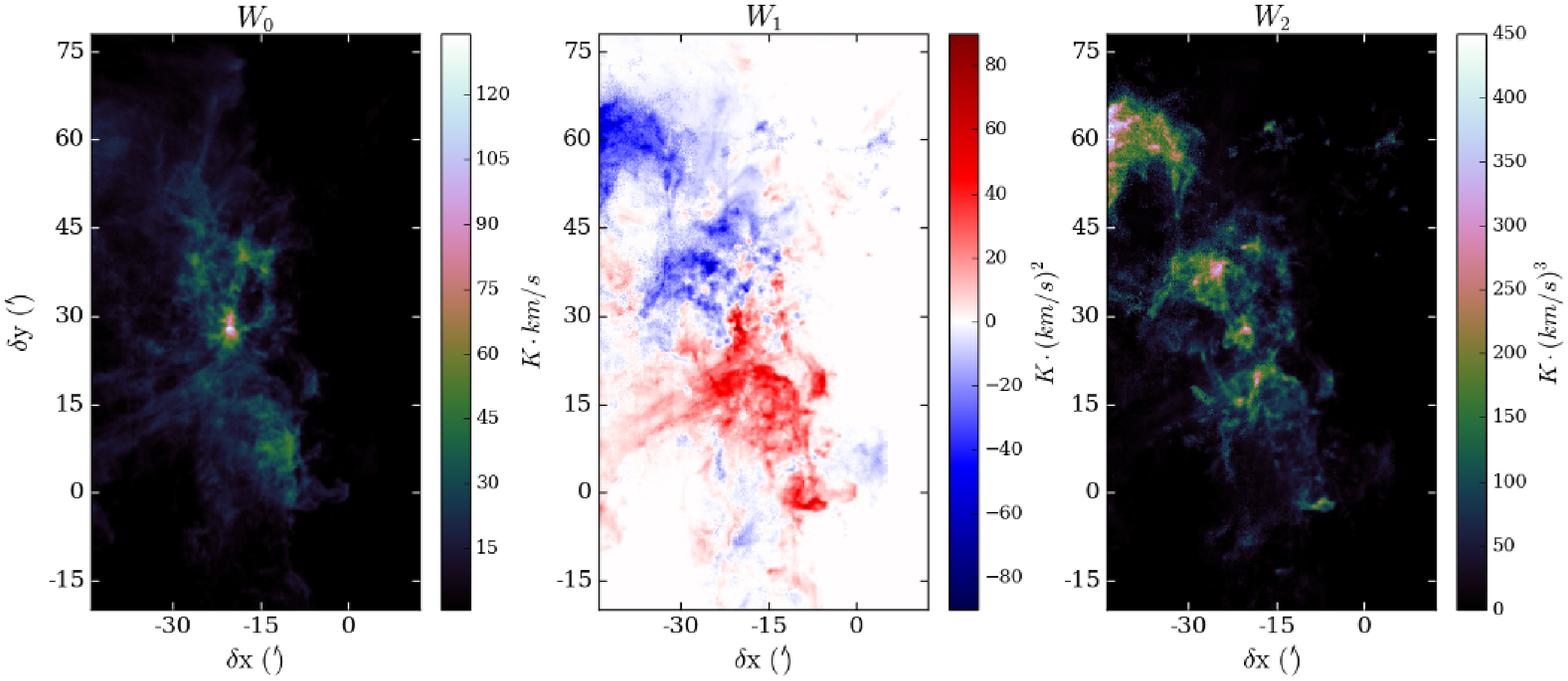} \\
    \includegraphics[width=\linewidth]{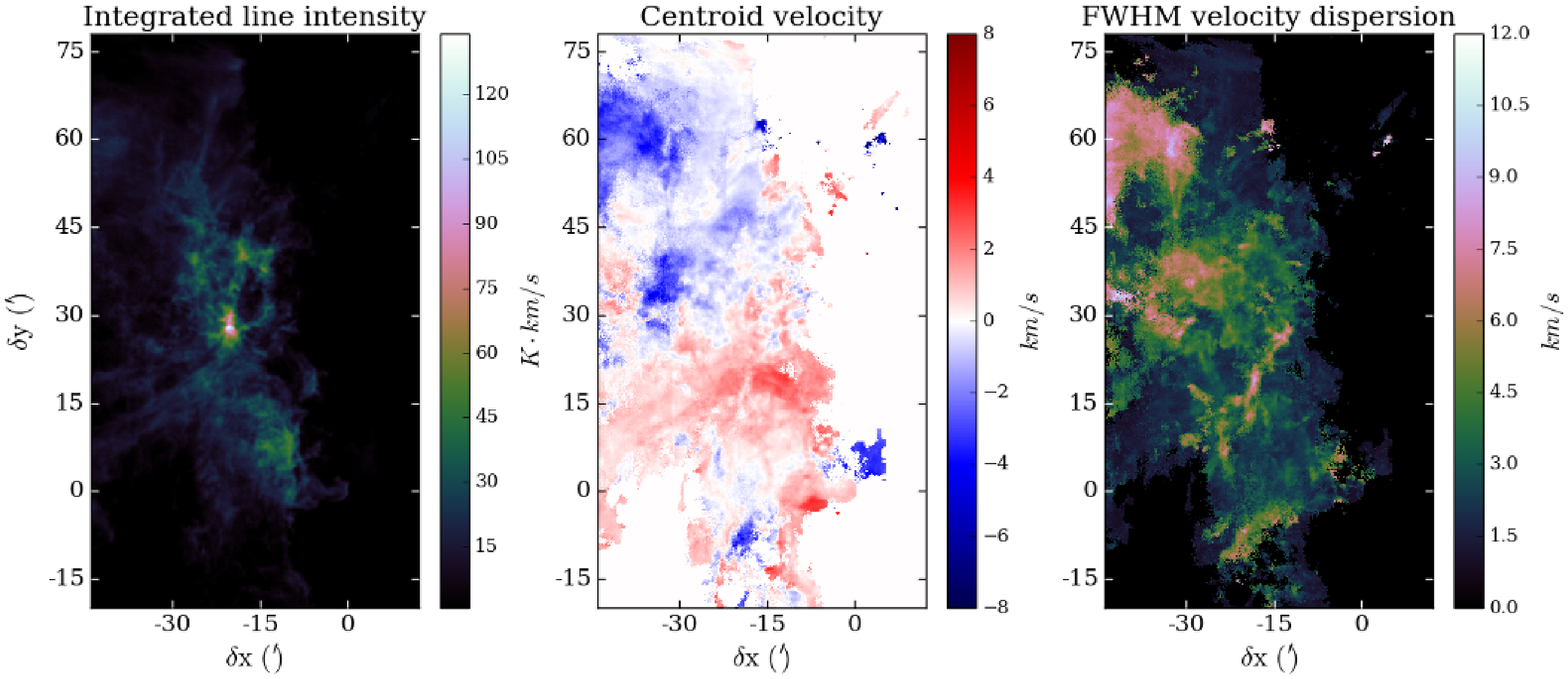}
    \caption{Top: maps of the $^{13}$CO($J=1-0$) field moments in the cloud's frame of reference. Bottom: maps of the physical quantities directly derived from each field, with the centroid velocity being simply $W_1/W_0$, and the FWHM velocity dispersion given by $2\sqrt{2\ln(2)}\sqrt{W_2/W_0}$ -- the normalization of the FWHM corresponds to that of a field with purely Gaussian line profiles.}
    \label{fig:moments}
  \end{figure*}}

\newcommand{\FigCompare}{%
  \begin{figure}
    \centering
    \includegraphics[width=\linewidth]{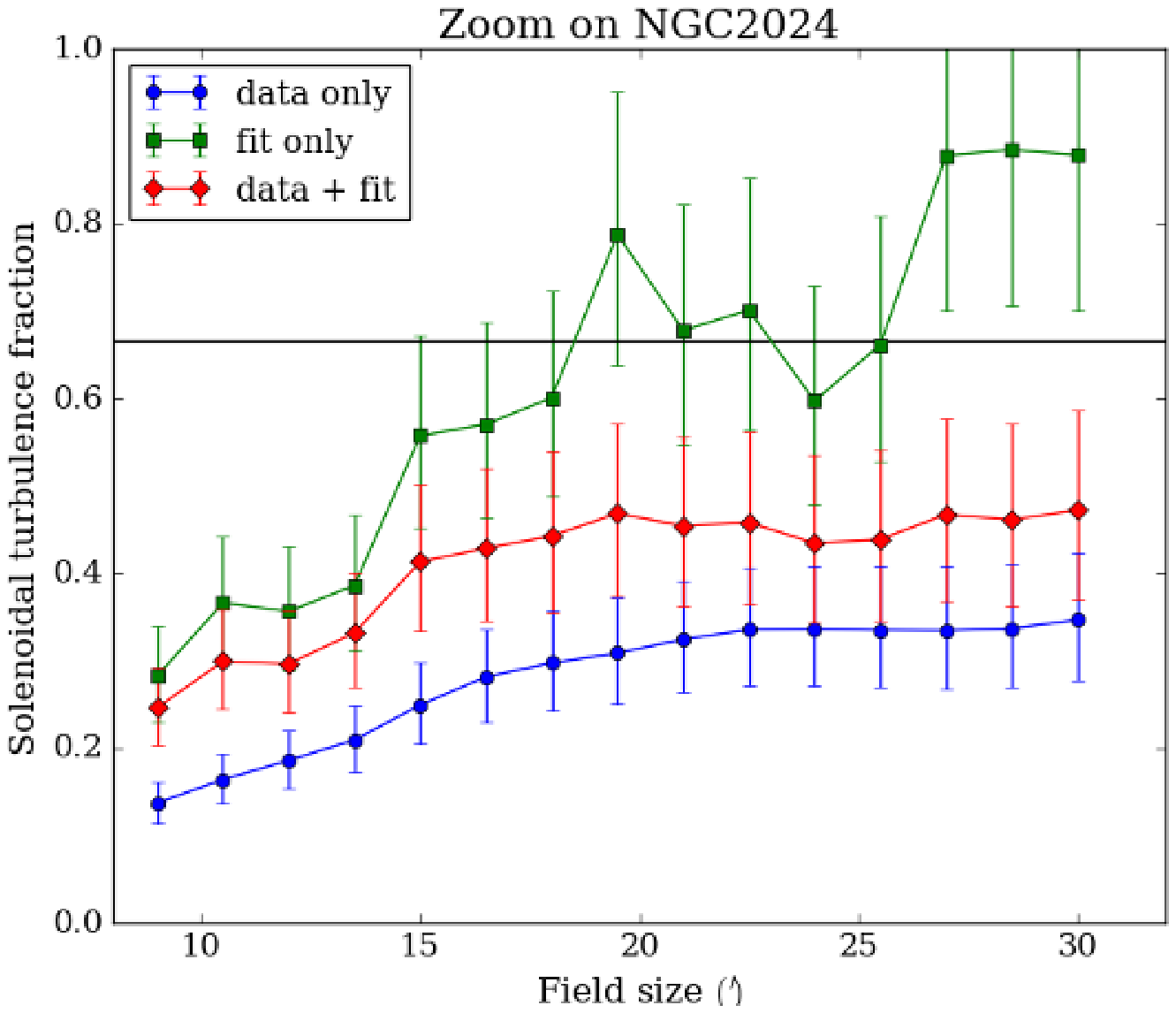}
    \caption{Comparison of the calculation results in the case of the zoom on NGC\,2024 with three different methods of computing the power spectra: either the power spectra resulting from Fourier-transformed data are used directly, with a linear interpolation between the points (data only, blue), or the result of the power law fit at high frequencies is extrapolated to low frequencies to provide a power law throughout the whole range of frequencies (fit only, green), or a composite version of the power spectrum is built using the raw data at low frequencies, and the fit result at high frequencies (data + fit, red). The error bars account for the observational noise as well as for the $g_{21}$ uncertainty. The horizontal line marks the equipartition limit.}
    \label{fig:compare}
  \end{figure}}

\newcommand{\FigGtwoone}{%
  \begin{figure}
    \centering
    \includegraphics[width=\linewidth]{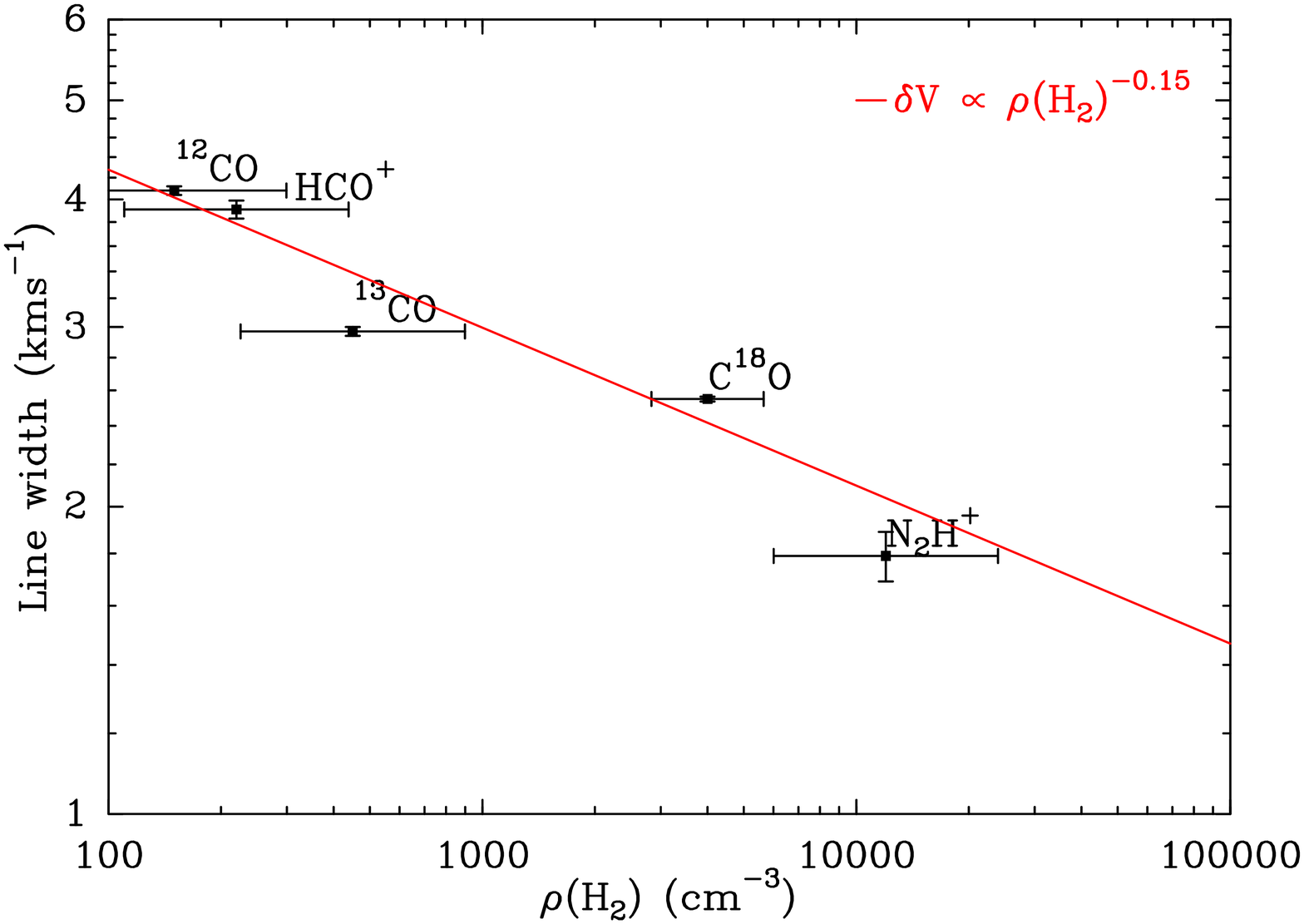}
    \caption{Variation of the Full Width at Half Maximum (FWHM) as a function of the gas density. Each point refers to the ($J=1-0$) spectral line of a different molecule. The red line shows the least squares fit, with a slope  $\alpha = -\epsilon = -0.15$. }
    \label{fig:g21}
  \end{figure}}

\newcommand{\TabMolecules}{
  \begin{table}
    \centering
    \caption{Spectral tracers used in this study, and observed with the IRAM-30m telescope. The linewidths are derived from these observations, the typical densities from these observations (ref. 1) or other works (ref. 2: \citet{hilyblant05}, ref. 3: \citet{kirk16}).}
    \begin{tabular}{ccccc}
      \hline
      \hline
	  Line & Frequency & FWHM & log($n($H$_2$)) & Ref. \\
	  ($J=1-0$) & (\GHz) & (km~s$^{-1}$) & (cm$^{-3}$) & \\ 
	  \hline
	  $^{12}$CO & 115.271202 & $4.08 \pm 0.04 $  &  $2.17 \pm 0.3\phantom{0}$ & 1 \\
	  HCO$^+$ & \phantom{0}89.188525 & $3.91 \pm 0.08 $  &   $2.34 \pm 0.3\phantom{0}$ & 1\\
	  $^{13}$CO & 110.201354 & $2.97 \pm 0.03 $  &  $2.65 \pm 0.3\phantom{0}$ & 1 \\
	  C$^{18}$O & 109.782173 & $2.55 \pm 0.01 $  &  $3.60 \pm 0.15$ & 2  \\
	  N$_2$H$^+$ & \phantom{0}93.173764 & $1.79 \pm 0.10 $   &  $4.10 \pm 0.3\phantom{0}$ & 3 \\
	  \hline
	\end{tabular}
    \label{tab:molecules}
  \end{table}}

\newcommand{\TabMach}{%
  \begin{table}
    \centering
    \caption{Typical values of the temperature and the Mach number distribution, for the two computed temperature maps. The $T_{\emr{exc}}$ distribution is truncated below 10\K, to avoid being biased by the large number of pixels at low $T_\emr{peak}$ in $^{12}$CO.}
    \begin{tabular}{ccccc}
      \hline
      \hline
      Value  & $T_{\emr{max}}$ & $T_{\emr{exc}}$ & $M_{\emr{max}}$ & $M_{\emr{exc}}$ \\
      \hline
      Most probable & 21.1 & 22.3 & 3.5 & 3.5 \\
      Mean          & 25.5 & 15.8 & 6.1 & 6.5 \\
      Median        & 23.4 & 15.6 & 4.8 & 5.0 \\
      \hline
    \end{tabular}
    \label{tab:mach}
  \end{table}}

\begin{document}

\title{Turbulence and star formation efficiency in molecular clouds: solenoidal versus compressive motions in Orion\,B\thanks{Based on observations carried out at the IRAM-30m single-dish telescope. IRAM is supported by INSU/CNRS (France), MPG (Germany) and IGN (Spain).}}
\titlerunning{Turbulence and star formation efficiency in Orion\,B}

\author{Jan H. Orkisz \inst{\ref{UGA},\ref{IRAM},\ref{LERMA-ENS}} %
  \and Jérôme Pety \inst{\ref{IRAM},\ref{LERMA-ENS}} %
  \and Maryvonne Gerin \inst{\ref{LERMA-ENS}} %
  \and Emeric Bron \inst{\ref{LERMA-M},\ref{ICMM}} %
  \and Viviana V. Guzm\'an \inst{\ref{CFA},\ref{JAO}} %
  \and Sébastien Bardeau \inst{\ref{IRAM}} %
  \and Javier R. Goicoechea  \inst{\ref{ICMM}} %
  \and Pierre Gratier \inst{\ref{LAB}} %
  \and Franck Le Petit \inst{\ref{LERMA-M}} %
  \and François Levrier \inst{\ref{LERMA-ENS}} %
  \and Harvey Liszt \inst{\ref{NRAO}} %
  \and Karin \"Oberg \inst{\ref{CFA}} %
  \and Nicolas Peretto \inst{\ref{UC}} %
  \and Evelyne Roueff  \inst{\ref{LERMA-M}} %
  \and Albrecht Sievers \inst{\ref{IRAM-SPAIN}} %
  \and Pascal Tremblin \inst{\ref{CEA}} %
}

\institute{%
  Univ. Grenoble Alpes, IRAM, F-38000 Grenoble, France \label{UGA} %
  \and IRAM, 300 rue de la Piscine, F-38406 Saint Martin d'Hères,
  France \label{IRAM} %
  \and LERMA, Observatoire de Paris, PSL Research University, CNRS, Sorbonne Universités, UPMC Univ. Paris 06, École normale supérieure, F-75005, Paris, France \label{LERMA-ENS} %
  \and LERMA, Observatoire de Paris, PSL Research University, CNRS, Sorbonne Universités, UPMC Univ. Paris 06, F-92190, Meudon, France \label{LERMA-M} %
  \and Laboratoire d'astrophysique de Bordeaux, Univ. Bordeaux, CNRS, B18N, allée Geoffroy Saint-Hilaire, 33615 Pessac, France \label{LAB} %
  \and Harvard-Smithsonian Center for Astrophysics, 60 Garden Street,
  Cambridge, MA, 02138, USA \label{CFA} %
  \and Joint ALMA Observatory (JAO), Alonso de Cordova 3107 Vitacura, Santiago de Chile, Chile \label{JAO}
  \and ICMM, Consejo Superior de Investigaciones Cientificas (CSIC).
  E-28049. Madrid, Spain \label{ICMM} %
  \and National Radio Astronomy Observatory, 520 Edgemont Road,
  Charlottesville, VA, 22903, USA \label{NRAO} %  
  \and School of Physics and Astronomy, Cardiff University, Queen's
  buildings, Cardiff CF24 3AA, UK \label{UC} %
  \and IRAM, Avenida Divina Pastora, 7, Núcleo Central, E-18012 Granada,
  España \label{IRAM-SPAIN}%
  \and Maison de la Simulation, CEA-CNRS-INRIA-UPS-UVSQ, USR 3441, Centre d'étude
de Saclay, F-91191
  Gif-Sur-Yvette, France \label{CEA} %
}

\date{}

\abstract%
{The nature of turbulence in molecular clouds is one of the key parameters that control star formation efficiency: compressive motions, as opposed to solenoidal motions, can trigger the collapse of cores, or mark the expansion of \Hii{} regions.}%
{We try to observationally derive the fractions of momentum density ($\rho v$) contained in the solenoidal and compressive modes of turbulence in the Orion\,B molecular cloud and relate these fractions to the star formation efficiency in the cloud.}%
{The implementation of a statistical method developed by Brunt \& Federrath (2014), applied to a $^{13}$CO($J=1-0$) datacube obtained with the IRAM-30m telescope, allows us to retrieve 3-dimensional quantities from the projected quantities provided by the observations, yielding an estimate of the compressive versus solenoidal ratio in various regions of the cloud.}%
{Despite the Orion\,B molecular cloud being highly supersonic (mean Mach number $\sim 6$), the fractions of motion in each mode diverge significantly from equipartition. The cloud's motions are on average mostly solenoidal (excess $> 8\%$ with respect to equipartition), which is consistent with its low star formation rate. On the other hand, the motions around the main star-forming regions (NGC\,2023 and NGC\,2024) prove to be strongly compressive.}%
{We have successfully applied to observational data a method that was so far only tested on simulations, and have shown that there can be a strong intra-cloud variability of the compressive and solenoidal fractions, these fractions being in turn related to the star formation efficiency. This opens a new possibility for star-formation diagnostics in galactic molecular clouds.}

\keywords{Turbulence - Methods: statistical - ISM: clouds - ISM: kinematics and dynamics - Radio lines: ISM - ISM: individual object: Orion\,B}

\maketitle{}

\section{Introduction}

The evolution of molecular clouds is controlled by a complex interplay of large-scale phenomena and microphysics: chemistry and interaction of the matter with the surrounding far-UV and cosmic-ray radiation control the thermodynamic state of the gas and its coupling to the magnetic field. The medium is highly turbulent, with Reynolds numbers reaching $10^7$ and magnetic Reynolds numbers reaching $10^4$ \citep{draine11}. Magneto-hydrodynamic (MHD) turbulence is one of the main counter-actions to gravity \citep{hennebelle12,federrath12,padoan14}, as well as the major mechanism that shapes the clouds: their fractal geometry is related to the properties of their turbulent velocity field \citep{pety00,federrath09}. The dissipation time scale for the turbulent energy of a molecular cloud is shorter \citep[$\sim 1\Myr$,][]{maclow99} than the age of such clouds \citep[$\sim 20-30\Myr$,][]{larson81}. Hence, a continuous energy injection must exist \citep[and references therein]{hennebelle12}. The proposed injection mechanisms may be either external, for instance Galactic shear or nearby supernovae explosions \citep{kim15}, or internal, like the expansion of \Hii{} regions and molecular outflows of the recently formed stars \citep{hennebelle12}. The nature of turbulence, and notably its solenoidal or compressive forcing, also plays a key role in the star formation efficiency (hereafter, SFE) of molecular clouds \citep{federrath12,federrath13}. In particular, compressive motions bear the mark of various phenomena related more or less directly to the formation of stars: infall on filaments, collapsing dense cores, expansion around young stars... As a result, we can expect that a more ``compressive'' cloud is a more ``active'' cloud, probably more likely to be forming stars, as proposed by \citet{federrath12}.
\\

In this work, we propose to measure for the first time (to our knowledge) the relative fractions of momentum in the solenoidal and compressible modes of turbulence in a molecular cloud, following a method devised and tested on numerical simulations by \citet{brunt10} and \citet{brunt14}. Our goal is to obtain a quantitative estimation of these fractions from observational data, and to compare their ratio with the star formation efficiency, derived from independent data. We investigate whether different fractions of compressive and solenoidal motions might provide a diagnostic for the variation of the star formation efficiency among molecular clouds. A different proportion of compressive forcing might be the reason why some molecular clouds form stars at a high rate while others do not \citep{federrath10,federrath12,renaud14}. 

Our object of study is a large region of a nearby giant molecular cloud, namely the south-western edge of the Orion\,B cloud (Barnard\,33 or Lynds\,1630). Orion\,B is relatively close to us, at a typical distance of $\sim 400\pc$ \citep{menten07,schlafly14}, so that a spatial resolution of $25''$ corresponds to 0.05\,pc or $10^4$\,AU in the cloud. The total mass of Orion\,B is estimated to be $7 \times 10^4\Msun$ \citep{lombardi14}, and the average incident FUV radiation field is $G_0 \sim 45$ \citep{pety16}. Orion\,B is located in the Orion GMC complex \citep{kramer96,ripple13}, east of the famous Orion\,Belt. Alnitak, the eastmost of the three belt stars shines in the foreground of the cloud. The south-western edge of the cloud represents an ideal laboratory to study star formation, and features several remarkable regions. First, the cloud is illuminated by the massive star $\sigma$\,Ori that creates an \Hii{} region, the emission nebula IC\,434, bounded on its eastern side by an ionization front. Silhouetted against this bright background, a dark cloud can be seen: the famous Horsehead nebula. HD\,38087 also creates a small \textsc{Hii} region, IC\,435. Still embedded in Orion\,B, the star-forming region NGC\,2024, known as the Flame Nebula, hosts several massive O-type young stellar objects, which have created compact \Hii{} regions inside the cloud. NGC\,2024 lies just east of the Alnitak star, and is crossed by a filament which is seen in absorption in visible light, and in emission in the radio range. The reflection nebula NGC\,2023 is a quieter counterpart of NGC\,2024, hosting young B-type stars. It lies North-East of the Horsehead nebula. The rest of the cloud contains extended and quieter areas with strong filamentary structures (Fig.~\ref{fig:cloud}). This area has been extensively observed in the 3\,mm range with the IRAM-30m telescope (PI: J. Pety). This survey has led to a series of papers (\citet{liszt16}; \citet{pety16}; \citet{gratier16}) to which this article belongs.
\\

The paper is organised as follows. We briefly describe in Section 2 the observations by the ORION-B (Oustanding Radio Imaging of OrioN B) collaboration, and the data we use here. In Section 3, we present the concepts and equations of the statistical method and the details of its implementation, from noise filtering to the computation of power spectra and the estimation of velocity-density correlations. The results are described in Section 4, and discussed in Section 5 with a special emphasis on the relation of the turbulence properties with the star formation efficiency in Orion\,B. In the Appendix, we present our computation of the Mach number map in the cloud.

\section{Observations}
\subsection{The Orion\,B project dataset}
The Orion\,B project (PI: J. Pety) has already mapped with the IRAM-30m telescope the south-western edge of the Orion\,B molecular cloud over a field of view of 1.5 square degrees in the full frequency range from 84 to 116\GHz{} at 200\kHz{} spectral resolution \citep{pety16}. The red rectangle in Fig.~\ref{fig:cloud} shows the field of view observed up to now, over an H$_2$ column density map produced by the \emph{Herschel} Gould Belt Survey consortium~\citep{andre10,schneider13}. The mapped area covers $56 \times 98$ arcminutes (about $6 \times 11$ parsecs at the assumed distance of Orion\,B, 400\pc{}) in size.

So far, the observations provided about 250\,000 spectra over a 32\GHz{} bandwidth, yielding a position-position-frequency cube of $370 \times 650 \times 160000$ pixels, each pixel covering $9'' \times 9'' \times 0.5$\kms{} (at Nyquist sampling). The reduced dataset amounts to 100 GB of data. The data reduction is described in details in \citet{pety16}. 

\FigCloud{} %

This dataset is unique for its extensive coverage, both spatially and spectrally, at a typical resolution of $25''$ and a median noise of 0.1\K{} $[T_{\emr{mb}}]$ per channel over a 32\GHz{} spectral range. From the spatial point of view, the survey gives us access to a large range of scales in the molecular cloud, from 50\mpc{} to 10\pc{}. For the analysis of turbulence properties (see Sect.~\ref{sec:turb}), it means that we are able to study a large fraction of the inertial range of the turbulence, with a potential view on the injection scale. The dissipation scale, on the other hand, is of the order of a milli-parsec \citep{hennebelle12,miville16}, and, at a distance of 400\pc{}, is only accessible using millimetre interferometers, and out of reach for the IRAM-30m telescope.

From the spectral point of view, having such a large bandwidth observed in one go allowed us to image over 20 chemical species \citep{pety16}, among which those listed in Tab.~\ref{tab:molecules}. As opposed to several small-bandwidth mappings, the spectral lines in this survey are observed in the same conditions and are well inter-calibrated, which gives an unprecedented spectral accuracy for such a large field of view.

\subsection{The $^{13}$CO spectral data cube}

Most of the work presented here was performed on the $^{13}$CO($J=1-0$) datacube, which covers a velocity range of 40\kms{} centred around the source systemic velocity of 10.5\kms{} and a rest frequency of $110.201\,354\GHz$. The datacube presents an RMS noise of $\sigma = 0.17\K$, and a median signal-to-noise ratio of $T_\emr{peak}/\sigma = 7.9$.

The $^{13}$CO($J=1-0$) line was chosen because it offers one of the highest signal-to-noise ratios over the whole map, but it does not feature as much saturation as the $^{12}$CO($J=1-0$) line. Signal is present at a signal-to-noise ratio greater than 5 in the whole map, except in the \textsc{Hii} regions around $\sigma$ Ori and HD\,38087 where molecular gas is photodissociated. The brightest regions are the NGC\,2023 nebula, the center of the NGC\,2024 nebula, and its northern edge (Fig.~\ref{fig:moments}, left panel).

$^{13}$CO is a good tracer of molecular gas, from moderately diffuse and translucent regions ($\Av = 1-5\magn$) up to moderately dense and shielded gas ($10^{4}\pccm$, $\Av = 10\magn$). CO has a small dipole moment (0.11\,D), hence the rotational lines have low Einstein coefficients \citep[\eg{}][]{mangum15}. This leads to relatively easy collisional excitation, excitation temperatures approaching the kinetic temperature, and moderate line opacities except for the most abundant species, $^{12}$CO. The abundance ratio $^{13}$CO/$^{12}$CO is equal to $^{13}$C/$^{12}$C or about 1/60 when chemical fractionation reactions, which are limited to the most diffuse regions, are inefficient \citep{wilson94}. Therefore, the $^{13}$CO abundance relative to H$_2$ remains approximately constant at a level of $\sim 2 \times 10^{-6}$ \citep{dickman78} across most of the cloud volume. $^{13}$CO starts to be depleted on dust grains in cold cores, but these represent only a small fraction of the mass and a negligible fraction of the volume of the Orion\,B molecular cloud \citep{kirk16}.

Figure~\ref{fig:channelmaps} shows the $^{13}$CO($J=1-0$) signal integrated over 4 contiguous velocity ranges. It showcases the complexity of the spectral structure of the cloud, with prominent variations of the line profile with the position, which is a consequence of the strong turbulence at play in the molecular cloud. 

\FigChannelMaps{} %

Up to four spectral components appear along each line-of-sight within the field (Fig.~\ref{fig:spectra}). A main component is visible around 10\kms, and a secondary component at lower velocity (about 5\kms). Sometimes an extra component at higher velocity (about 14\kms), or secondary peaks around 5 and 10\kms appear too. The first two components are the most significant ones at the scale of the whole cloud, being the only ones visible in the mean spectrum of $^{13}$CO($J=1-0$), and have average velocities of 9.7\kms{} and 4.9\kms{} respectively. All the components, despite being quite distinct on the spectral axis, are however thought to be part of the Orion\,B cloud \citep[see discussion in][]{pety16}.

\FigSpectra{} %

\section{Deriving the relative fraction of solenoidal motions from a position-position-velocity cube}

In order to measure the fraction of the solenoidal and compressive turbulence modes, we apply the method developed by \citet{brunt10} and \citet{brunt14}. In this section, we first recall the method, its assumptions, and the way we implemented it.
     
\subsection{Description of the method}
\subsubsection{Principles and assumptions}

The key point of this method is the fact that the objects we observe are fundamentally 3-dimensional (\eg{} a molecular cloud), but the observer only has access to a 2-dimensional projection along the line of sight of that object. \citet{brunt10} developed a method to retrieve properties of the 3-dimensional object that we are interested in, and which corresponds to a 3-dimensional field $F_\emr{3D}$, via the properties of the 2-dimensional observational $F_\emr{2D}$, which is a projection of $F_\emr{3D}$ along the $z$ axis. To achieve this, they use the fact that the Fourier transform $\tilde{F}_\emr{2D}$ of the 2-dimensional field is proportional to the $k_z = 0$ cut through the Fourier transform $\tilde{F}_\emr{3D}$ of the 3-dimensional field. In short, $\tilde{F}_\emr{2D}(k_x,k_y) \propto \tilde{F}_\emr{3D}(k_x,k_y,k_z = 0)$. If these fields are isotropic, \ie{} if they are functions of $k = |\mathbf{k}|$ alone, with $\mathbf{k} = (k_x,k_y,k_z)$ the wave vector and $k$ the wave number, the 2-dimensional field allows us to reconstruct average properties of the 3-dimensional field thanks to symmetry arguments.

The \citet{brunt14} method was developed as an application of the \citet{brunt10} method to the case of vector fields. For a vector field such as a velocity or momentum field, the dimensionality reduction due to the projection is made worse by the fact that only one component of the vector (the line-of-sight one) can be measured, thanks to the Doppler effect. In that case, the next main tool to retrieve 3-dimensional properties is the Helmholtz theorem, which allows to decompose any vector field in its divergence-free (solenoidal) and curl-free (compressive) components, $\mathbf{F}_\perp$ and $\mathbf{F}_{||}$. These components are related via a \emph{local} orthogonality in Fourier space, $\tilde{\mathbf{F}}_\perp(\mathbf{k}) \perp \tilde{\mathbf{F}}_{||}(\mathbf{k})$. Solenoidal modes can be pictured as the modes of a turbulent incompressible field, made of vertices and eddies. On the other hand, compressive modes, made of compression and expansion motions, are more likely to be generated by phenomena linked to star formation.

The application of these methods implies several requirements on the studied dataset. As mentioned earlier, the statistical isotropy of the cloud is the first necessary point, and enables the use of 2-dimensional averages as a means to estimate 3-dimensional properties. It means that the method cannot be applied to individual filaments, or to clouds where a strong anisotropy is suspected, \eg{} due to the presence of a strong magnetic field at low Mach numbers.

Second, the field is required to go smoothly to zero on its borders. This property is needed to ensure that the decomposition of the field is unique, since the Helmholtz decomposition is, in theory, defined up to a vector constant. It is also a necessary condition for good behaviours of the Fourier transform, since actual observational data are not periodic fields, unlike hydro-dynamical simulations \citep[see discussion in][]{brunt10}. This implies that the studied field should be bounded in space, like for example a gravitationally bound cloud. In the case where the signal extends up to the edge of the observed field, the dataset has to be apodized. 

Finally, from a practical viewpoint, \citet{brunt10} have shown that their method works best for fields with power spectra that are not too steep. Steep power spectra give measurements that are very sensitive to the low spatial frequencies, which are usually uncertain due to poor statistics.

The compliance of our dataset with these requirements is discussed in detail in Sect.~\ref{sec:compliance}.

\subsubsection{Equations and notations}

The studied quantity is the momentum density field (hereafter ``momentum''), $\mathbf{p} = \rho \mathbf{v}$, with $\rho$ and $\mathbf{v}$ the volume density and the velocity.

The 3-dimensional quantity we infer is
\begin{equation}
R = \sigma_{p_\perp}^2 / \sigma_p^2,
\end{equation}
the ratio of the variance of the transverse (solenoidal) momentum to the variance of the total momentum. For short, $R$ will be referred to as the ``solenoidal fraction'' in the rest of this paper.

According to \citet{brunt14}, at hypersonic Mach numbers ($M = v/c_{\emr{sound}} > 5$) the solenoidal fraction does not depend any more on the type of forcing, but instead converges towards $R \sim 2/3$. \citet{brunt14} note that this behaviour is different from what is observed by \citet{federrath11}, where the solenoidal fraction converges to different values depending on the type of forcing, but this is due to the fact that \citet{brunt14} consider the \emph{momentum} density field, while \citet{federrath11} describe the \emph{energy} density field.

This value of $R \sim 2/3$ can be simply explained in terms of equipartition of momentum between the compressive and solenoidal modes \citep[see, \eg{}][]{federrath08}. A value of $R$ lower than 2/3 therefore means that there is more momentum in the compressive modes of the flow, and that the cloud is thus more likely to form stars. The influence of the Mach number is further discussed in Sect.~\ref{sec:mach}.

The available observables are a position-position-velocity  cube, its velocity-weighted moments and the power spectra of these moments. We make two major assumptions about this datacube, namely that the $^{13}$CO($J=1-0$) line is optically thin and that its emissivity only depends on the $^{13}$CO volume density. These assumptions are true within less than 20\%, with the exception of a few lines of sight towards the center of NGC\,2024, which are more saturated, but represent about 2\% of the whole field. Under these assumptions, the position-position-velocity cube can be seen as a density-weighted velocity field: the spectrum obtained for each line of sight results from the projection of the emission of the matter present along this line of sight, and moving at various velocities. 

The useful moments are the zero-th, first and second order moments of the momentum field, $W_0$, $W_1$ and $W_2$, which are defined as follows in \citet{brunt14}:
\begin{equation}
  W_0 = \!\int\!\!I(v) \mathrm{d}v, \quad W_1 = \!\int\!\! v I(v) \mathrm{d}v, \quad W_2 = \!\int\!\! v^2 I(v) \mathrm{d}v.
\end{equation}

The spectral line intensity, $I(v)$, may have contributions from various positions along the line of sight. Given our assumptions on emissivity, and assuming that the natural linewidth is negligible compared to the overall velocity dispersion, we can describe these moments in an alternative way:
\begin{equation}
  W_0 \propto \!\int\!\!\rho(z) \mathrm{d}z, \quad W_1 \propto \!\int\!\! v(z) \rho(z) \mathrm{d}z, \quad W_2 \propto \!\int\!\! v(z)^2 \rho(z) \mathrm{d}z.
\end{equation}
The solenoidal fraction can be written in terms of these observational quantities as follows (see \citet{brunt14} for details):
\begin{equation}
  \label{eq:R}
  R \approx \left[\frac{\langle W_1^2 \rangle}{\langle W_0^2\rangle}\right]\! \left[\frac{\langle W_0^2 \rangle / {\langle W_0 \rangle}^2}{1 + A (\langle W_0^2 \rangle / {\langle W_0 \rangle}^2-1)}\right]\! \left[ g_{21}\frac{\langle W_2\rangle}{\langle W_0\rangle}\right]^{-1}\!\!\! B.
\end{equation}
The $A$ and $B$ factors are functions of the power spectra of $W_0$ and $W_1$.
\begin{equation}
  \label{eq:AB}
 A = \frac{\left(\sum_{k_x}\sum_{k_y}\sum_{k_z}f(k)\right)-f(0)}{\left(\sum_{k_x}\sum_{k_y}f(k)\right)-f(0)}, \enskip B = \frac{\sum_{k_x}\sum_{k_y}\sum_{k_z}f_{\perp}(k)\frac{k_x^2+k_y^2}{k^2}}{\sum_{k_x}\sum_{k_y}f_{\perp}(k)},
\end{equation}
where $f(k)$  and $f_{\perp}(k)$ are the angular averages of the power spectra $\tilde{W}_0(k_x,k_y)$ and $\tilde{W}_1(k_x,k_y)$, respectively.

\citet{brunt14} introduce a statistical correction factor, $g_{21}$, of order unity, which measures the correlations between the variations of the density and the velocity fields, and is defined as
\begin{equation}
 g_{21} = \frac{\langle \rho^2 v^2 \rangle / \langle \rho^2 \rangle}{\langle \rho v^2 \rangle / \langle \rho \rangle}.
\end{equation}
\citet{brunt14} show that this may be written as
\begin{equation}
 \label{eq:g21}
 g_{21} = \left\langle\left(\frac{\rho}{\rho_0}\right)^2\right\rangle^{-\epsilon}.
\end{equation}
The 3-dimensional variance of the volume density, $\langle (\rho/\rho_0)^2\rangle$, can be derived using the \cite{brunt10} method. The exponent $\epsilon$ is a small, positive constant, which can be obtained as the exponent of the $\sigma_v^2$ vs. $\rho$ power law, \ie{} the typical velocity dispersion in the cloud as a function of volume density. If the density and velocity fields were uncorrelated, $g_{21}$ would be equal to 1.

\subsection{Implementation}

Actual data suffer from several limitations that need to be dealt with in order to apply the method described previously. The field of view, the angular resolution, and the sensitivity are limited. This section describes how these issues were dealt with.

\subsubsection{Noise filtering}
\label{sec:noisefiltering}
Computation of line moments is sensitive to noise in the line wings. It is well-known that masking the position-position-velocity cube where signal stays undetected improves the determination of the centroid velocity and linewidth. To define the mask containing the pixels detected at high significance, these pixels are first grouped into continuous brightness islands that are made of neighbouring pixels in the position-position-velocity space, whose signal-to-noise ratio is larger than 2. The noise level, $\sigma$, is measured outside the studied velocity range $(-0.5,\, 20.5\kms)$. The list of islands is then sorted by decreasing total flux. The first island contains
about 97.8 \% of the total signal. The following ones are small signal clumps that are spatially or spectrally isolated from this main block, and after about a few hundred islands we are left with single-cell islands which are just noise peaks. 

While it is easy to visually assess that the first few islands correspond to genuine signal, it is more complex to determine the transition to pure noise, as a significant fraction of the total line flux could be hidden in pixels of faint brightness, at low signal-to-noise ratio. We thus studied the influence of the number of islands used on $R$, the solenoidal fraction in the studied cube. This influence is modest, mainly because over 97\% of the signal is located in the first island. Using up to about 80 islands yields a very stable value of $R$ with less than 0.1\% variation. A steep increase of $R$ is observed when we enter the noisy domain. We thus used the 80 brightest islands for all other calculations.

\subsubsection{Moment computation}
After selecting the signal islands in the position-position-velocity cube, the moments are integrated from $-0.5$ to 20.5\kms{}. The calculations have to be performed in the center-of-mass frame of reference of the cloud, which implies determining the centroid velocity of the cloud in the LSR frame. This center-of-mass velocity is simply given by
\begin{equation}
V_\emr{c} = \frac{\langle W_1^\emr{obs} \rangle}{\langle W_0 \rangle},
\end{equation}
where $W_1^\emr{obs}$ is the first moment field in the observer's frame of reference. $W_0$, on the other hand, is not velocity-weighted, and therefore does not depend on the frame of reference. For the observed field of view, we obtain $V_\emr{c} = 9.16 \pm 0.90$\kms{}.

The velocity scale in the observer's frame of reference is shifted by $V_\emr{c}$, before computing $W_1$ and $W_2$ in the center-of-mass frame of reference:
\begin{equation}
W_{1}\! = \!\!\!\int\!\! (v_\emr{obs}-V_\emr{c}) \cdot I(v_\emr{obs}) \mathrm{d}v_\emr{obs}, \enskip W_{2} \!= \!\!\!\int\!\! (v_\emr{obs}-V_\emr{c})^2\cdot I(v_\emr{obs}) \mathrm{d}v_\emr{obs}.
\end{equation}
The resulting fields are shown in Fig.~\ref{fig:moments}.

\FigMoments{}

\subsubsection{Apodization}

Computing the power spectra of the $W_0$ and $W_1$ fields requires to take the Fourier transform of these fields. Although calculating the FFT of 2-dimensional fields is an easy task, several numerical artefacts must be taken care of. In particular, the observed area does not reach the edges of the Orion\,B molecular cloud in all directions as illustrated in Fig.\ref{fig:cloud}. This sharp truncation of the $^{13}$CO emission will create artefacts in the Fourier transform, due to the convolution of the true Fourier spectrum by a sinus cardinal function that oscillates at high frequencies. Apodizing the field is required to avoid this behaviour. We have chosen to multiply the intensity by $1 - \cos(\pi x/w)$, where $x$ is the pixel coordinate and $w$ the apodization width. This function goes from 0 to 1 over $w$ pixels. This apodization function is used in \citet{martin15}. We keep the region affected by apodization as small as possible to minimize signal alteration. An apodization width of about 5 \% of the smallest dimension of the field (\ie{} roughly 25 pixels) was the smallest value that efficiently smoothed out the high-frequency artefacts. This is consistent as well with the width determined in \citet{martin15}.

Once the field is apodized, it must be made square to follow the isotropy requirements of the method. The square was built by padding with zeros the right side (west) of the observed area, as this is the location of the \Hii{} region associated with $\sigma$\,Ori, \ie{} no signal is detected past the western edge of the field of view. After this, the Fourier transform is calculated using the FFT implementation of Numpy 1.8 \citep{CT65}.

Apodization is a linear filter of the data, and thus has effects on the power spectrum at all frequencies. While apodization allows us to ``clean'' the spectrum at high spatial frequencies, it also alters the spectrum at the lowest frequencies. For example, $f(0)$ -- the value of the power spectrum at spatial frequency $k=0$ -- is directly proportional to the spatial integral of $W_0^2$, therefore losing some signal due to the apodization will reduce the value of $f(0)$. The method chosen to keep the good parts of the apodized and non-apodized spectra was the following: we first computed the FFT of both the apodized and non-apodized fields, respectively $\tilde{W}_\emr{ap}$ and $\tilde{W}_\emr{nap}$, then mixed them smoothly around $k = 0$ using a narrow (about 10 Fourier-space pixels) 2-dimensional Gaussian $G_\emr{mix}(k)$. The resulting Fourier field is
\begin{equation}
  \label{eq:fourier}
 \tilde{W}_\emr{final} = G_\emr{mix}(k)\cdot\tilde{W}_\emr{nap}(\vec{k}) + \left(1- G_\emr{mix}(k)\right)\cdot\tilde{W}_\emr{ap}(\vec{k})
\end{equation}
The power spectrum thus behaves as the non-apodized spectrum at low $k$, keeping the correct value of the field integral $f(0)$, and as the apodized spectrum at high $k$, free of the spectral parasites created by the sharp edges of the map.

\subsubsection{Power spectra computation and fit}
\label{sec:powerspectra}
The apodized and corrected FFT of the field needs to be transformed into an angle-averaged power spectrum $f(k)$ or $f_\perp(k)$. It is simply done by binning the modulus of the spatial frequencies, and averaging the points found in these radial bins. The resulting discrete function can then be linearly interpolated into a continuous function. A critical element to this exercise resides in the sampling of the spatial frequency axis. On the one hand, the resulting angle-averaged spectrum should be as detailed as possible, but on the other hand, a larger number of bins can lead to empty bins, containing no sampled points at all. As a result, we used a number of bins of $S/1.45$, with $S$ the size in pixels of the square field, so that the size of a bin in the Fourier space corresponds to slightly more than the length of the diagonal of pixels in the Fourier space.

Additional observational constraints (noise, beam shape, etc.) affect the determination of the power spectrum. Following \citet{martin15}, the power spectra are fitted with a power law, modified to take into account the single-dish beam and the noise. In our case, the beam is modelled as the Fourier transform of a Gaussian of FWHM equal to the cube resolution, \ie{} $23.5''$. This corresponds to about 2.61 pixels. The convolution in the image space corresponds to a multiplication in the Fourier space that mostly affects the high spatial frequencies.

The noise is not a Gaussian white noise, because of inter-pixel correlations and systematics. We therefore use the power spectra of 30 signal-free channels and average them to obtain a template of the noise power spectrum. In this case, we use the fully noisy data cube, not the 80 first signal islands, in order to have the same spatial correlations in noise for each channel (with or without signal), which would not be the case with a masked data cube. The noise template intends to reproduce a systematic behaviour, but we can only use a finite number of channels with random noise. The noise template is therefore smoothed to make this systematic pattern stand out more. Given that $W_1$ is, just like $W_0$, a linear combination of the channel maps, the same noise template is used for both power spectra.

\FigWzero

The fit is performed in the log($k$)-log($f$) space, so that the straight line of the power law stands out more. The fitting function is therefore the logarithm of
\begin{equation}
  10^{\left(a\cdot\log(k) + b\right)}\cdot \tilde{G}_\emr{beam}(k)^2 + N\cdot noise(k)
\end{equation}
where $G_\emr{beam}$ is the Gaussian beam, $\tilde{G}_\emr{beam}$ is its (Gaussian) Fourier transform, and $noise$ is our noise template. The fitted parameters are $a$, $b$ and $N$. During the fit, the data are weighted by the inverse of the variances obtained in each bin when computing the power spectrum. 

The choice of fitting the power spectrum with a modified power law implies that the underlying physical processes should produce a power law. This is indeed the case for the inertial range of scales in Kolmogorov turbulence, and can be applied as well to Burgers turbulence \citep[see, \eg{}][]{federrath13b}. However, the power spectrum of turbulence starts to deviate from a power law at scales where the energy is injected (low spatial frequencies) or dissipated (high spatial frequencies). The power spectra computed from our dataset somewhat deviates from a power law at low spatial frequencies (see Fig.~\ref{fig:W0}). The power law can therefore only be fitted and then used above a given spatial frequency. The power-law range starts around $\sim 5.5'$. At high spatial frequencies, no deviations from the power law are detected (the fits render the observations very well). This means that either the dissipation scale happens at lower angular scale than the resolution of the observations or it is hidden in the noise.
\\

Once the fit has been performed, the final version of the power spectra is built using both the fit result and the observational angle-averaged power spectra, and used for further calculations. The final power spectrum is a pure power law (without the beam and noise) above the $1/5.5$ arcmin$^{-1}$ threshold, and is equal to the linearly interpolated angle-averaged power spectrum below this threshold. In particular, we enforce $f_\perp(0) = 0$, since we are working in the cloud's rest frame.

\subsection{Density-velocity correlations}
\label{sec:g21}

We used the information on the mean line profiles to estimate the slope $\epsilon$ of the relation between the velocity dispersion and the local density  (see Eq.~\ref{eq:g21}). Among the lines detected in the mean spectrum, we have selected lines with different spatial distributions \citep{pety16}. To trace the low density gas, we have selected $^{12}$CO($J=1-0$) and HCO$^+$($J=1-0$) as these lines present very extended emission and have moderate excitation requirements \citep{pety16,liszt16}. We have included  $^{13}$CO($J=1-0$) as our tracer of the bulk of the gas. The somewhat denser and more shielded gas is well traced by C$^{18}$O($J=1-0$), while we have selected N$_2$H$^+$($J=1-0$) for the dense cores. For these five species, we determined the FWHM by fitting a Gaussian line profile to the mean profile of the whole map. Only the 10\kms{} component, which is present for all five species, was used for this fit. For N$_2$H$^+$ we used the HFS fit method in GILDAS/CLASS\footnote{See \texttt{http://www.iram.fr/IRAMFR/GILDAS/} for more details on the GILDAS software.}, which makes use of the information on the hyperfine structure.

While these lines are emitted by gas over a wide range of densities, there is a minimum density under which the line is not detected, because of lack of excitation or because the molecule is not present in low density gas. It is this minimum density which corresponds to velocity dispersion of the line. To derive the densities associated with the line emission,  we adopted three different methods.

For the low density and extended emission tracers, we derived the volume density by comparing the minimum gas column density where the emission is detected and the resulting size of the emission regions. This leads to gas densities of a few hundred cm$^{-3}$ for $^{12}$CO($J=1-0$), HCO$^+$($J=1-0$) and  $^{13}$CO($J=1-0$). The emission of $^{12}$CO($J=1-0$) is dominated by the low density regions \citep{pety16}. The emission of HCO$^+$($J=1-0$) is also dominated by low to intermediate density regions, and comes from the weak excitation regime \citep{liszt16}. In both cases, opacity broadening is not very significant. We must however consider that the line widths for these tracers are upper bounds, due to this effect of opacity broadening. For these molecules, we are also limited by the sensitivity of the observations, so that the densities should be regarded as upper limits. \citet{hilyblant05} analysed the structure and kinematics of the Horsehead nebula and derived the density of the extended region traced by C$^{18}$O as $3 - 5 \times 10^{3}$~cm$^{-3}$. We have kept this value as the typical density traced by C$^{18}$O.  \citet{hilyblant05} have shown that higher density regions exist with densities significantly larger than 10$^4$~cm$^{-3}$. As very few pixels are detected in N$_2$H$^+$ in the region of the Horsehead nebula, we have used the catalogue of dense cores  identified by \citet{kirk16} in their SCUBA2 map of the Orion\,B complex. We have found 55 cores associated with N$_2$H$^+$($J=1-0$) emission. The densities have been derived using the extracted fluxes and effective radii, a uniform dust temperature of 20\K, and assuming spherical geometry for all cores. The mean density is 10$^{4.1}$~cm$^{-3}$ with a scatter of about a factor of two. The temperature of the N$_2$H$^+$ cores was not individually checked, but it is likely to be lower than 20\K. In turn, this implies an even higher density of N$_2$H$^+$. Therefore, the derived density should be regarded as a lower limit.

Table \ref{tab:molecules} presents the resulting data, which are illustrated in Fig.~\ref{fig:g21}. The slope $\alpha = -\epsilon$ is derived from a least square fit of the variation of the FWHM with the density. We derive $\epsilon = 0.15 \pm 0.03$. The possible systematics on the densities traced by $^{12}$CO($J=1-0$), HCO$^+$($J=1-0$) and N$_2$H$^+$($J=1-0$), as well as on the linewidths of $^{12}$CO($J=1-0$) and HCO$^+$($J=1-0$), all tend to make the power law steeper. Therefore, we keep this value of $\epsilon$ as an upper limit.

\TabMolecules{} 

\FigGtwoone{}

\subsection{Estimation of the uncertainties}
The computation of uncertainties implies computing the uncertainty of each element of Eq.~\ref{eq:R}. We start from the average RMS noise level in the data cube, 0.17\K{} $[T_{\emr{mb}}]$ . This allows us to compute the noise level for the $W_0$, $W_1$ and $W_2$ maps. The computation is straightforward compared to the computation of the uncertainty of the centroid velocity and linewidth because $W_1$, and $W_2$ are not normalized by $W_0$, \ie{} their noise distributions stay Gaussian, whatever the value of $W_0$.
Due to the velocity weighting, the absolute uncertainty increases significantly with the moment order. However, the relative uncertainties on $\langle W_0 \rangle$, $\langle W_1 \rangle$ and $\langle W_2 \rangle$ are similar, with values  ranging from about 10\% for the whole field to 5\% for the deepest zooms on NGC\,2023 and NGC\,2024 (see Fig.~\ref{fig:zoom}).
The uncertainties on the sums $A$ and $B$ were explicitly computed according to the error bars described in Sect.~\ref{sec:powerspectra}. The relative uncertainty on $A$ ranges from 24\% for the full field to 11 \% for the deepest zooms, and stays around 13\% for $B$.

For the overall relative uncertainty, one must not only take into account the errors of the individual terms, but also the correlation between the different variables. In our case, the different variables are strongly correlated, since they are all by-products of the same data cube. Therefore, we chose to use the average of the various relative errors as a rule-of-thumb estimate of the overall error. This approach yielded an overall relative error of 13\% for the whole field, and 8\% for the deepest zooms. We kept the highest value to allow for a safety margin.

This 13\% relative error $\Delta R/R$ corresponds to the median noise-to-signal ratio in the field, $\sigma/T_\emr{peak}$, which is also of the order of 13\% -- but testing with simulated noise whether this is a coincidence or not is out of the scope of this article. 

\section{Results}

In this section, we first briefly present the derivation of the Mach number in the cloud, then we compare the obtained power spectra with other results in molecular clouds, and finally, we give the results of our computation of the solenoidal fraction $R$ in the Orion\,B cloud.

\subsection{Mach number}
\label{sec:mach}
According to \citet{brunt14} and \citet{federrath11}, at hypersonic Mach numbers ($M > 5$) the ratio of solenoidal and compressive modes does not depend any more on the type of forcing. It is thus important to also derive the distribution of the Mach number to be able to interpret the results.

Using the maps of the sound speed derived from $^{12}$CO$(J=1-0)$ and dust temperature maps, two different estimations of the Mach number, $M_\emr{max}$ and $M_\emr{exc}$, were computed (see Appendix for details). Figures \ref{fig:mach:ima} and \ref{fig:mach:hist} show their spatial distributions and compare their histograms. The shapes of the histograms of Mach numbers computed with $T_\emr{max}$ and $T_\emr{exc}$ are very similar, and both show a large tail at high Mach number. Table \ref{tab:mach} lists several characteristic values of both distributions. The most probable value ($\sim3.5$) of the Mach number is much smaller than the mean or median values ($\sim6$), for both distributions.

\citet{schneider13} estimate the average Mach number to be of $\sim 8$, with approximately $30 - 40$\% error, deriving this value from \emph{Herschel} dust temperature and CO linewidth, and find out that Orion\,B has the highest Mach number of the set of studied clouds. Our results are compatible with this mean value, and they also provide the spatial and statistical distribution at good angular resolution. In particular, the Mach number is much smaller than the average in the star-forming regions NGC\,2023 and NGC\,2024, below the hypersonic regime.

\subsection{Power spectra}

When the whole field is considered, the fit yields an exponent $a_0 = -2.83 \pm 0.02$ for the $W_0$ field, and $a_1 = -2.50 \pm 0.07$ for the $W_1$ field. When zooming into specific regions of the cloud (NGC\,2023 and NGC\,2024, see Fig.~\ref{fig:zoom}), the values of these exponents range from $a_0 = -2.52 \pm 0.08$ (widest field) to $a_0 = -3.04 \pm 0.05$ (smallest field), and from $a_1 = -2.24 \pm 0.09$ to $a_1 = -2.81 \pm 0.06$. While the indices of $W_1$ power spectra are rarely reported, there are many values of spectral indices for integrated line intensity maps in the literature, and our $a_0$ values fall well in the range of spectral indices for observations of CO emission, dust emission, \textsc{Hi} emission and absorption compiled by \citet{hennebelle12}: the values range from -2.5 to -3.2, with most values around -2.7.

In order to have a meaningful result for the $A$ and $B$ coefficients, the power law slope of the power spectra must follow the two steepness requirements mentioned by \citet{brunt10}. On the one hand, the spectra should not be too steep, so that the weight of the low spatial frequencies, for which the available information is scarce, does not become too large in the $A$ and $B$ sums. On the other hand, the slopes should be steep enough to avoid divergence of these sums at large frequencies. A slope of $a = -3$ is at the limit between these two contradicting constraints, since the divergence of the 3-dimensional sum is only logarithmic. In our case, the sums are finite, due to the finite resolution of the observations, so that with slopes between -2.24 and -3.04, the sums do not grow too quickly and do not give too much weight to the low spatial frequencies.

\subsection{Turbulence mode ratio}
\label{sec:turb}
We have determined a relative error of about 13\% on the calculation of the ratio $R$ from the position-position-velocity cube. The correction factor $g_{21}$ is determined independently, and we assume a range of possible values for $g_{21}$, the lower limit being given by our calculations of Sect.~\ref{sec:g21}, and the upper limit resulting from the minimum estimate of $\epsilon \simeq 0.05$ according to \citet{brunt14}.

For the entire $^{13}$CO field, we obtain the following range of values:

\begin{equation}
  %R = \left(\frac{\sigma^2_\perp}{\sigma^2}\right)_{^{13}\emr{CO}} = 0.88^{+0.12}_{-0.30}  
%  0.72^{+0.14}_{-0.14}  < R_{^{13}\emr{CO}} <  1^{+0.0}_{-0.18}
  0.72^{+0.09}_{-0.09}  < R_{^{13}\emr{CO}} <  1^{+0.0}_{-0.09}
\end{equation}

In order to gain a deeper understanding of the dynamics at stake in the Orion\,B cloud, the method was also applied to several sub-regions of the $^{13}$CO map.

First, to check the reliability of the method and the homogeneity of the field, we applied a sliding square window, whose side is equal to the smallest dimension of the mapped area (Fig.~\ref{fig:slide}). This avoids zero-padding the studied field. The results show that even though the values are in general somewhat lower for these sub-fields than for the whole $^{13}$CO field, they remain marginally compatible with this result, within the estimated uncertainties.

\FigSlide{}

Second, we searched for systematic variations of the fraction of solenoidal modes when zooming into specific regions of the map.  In particular, signs of the solenoidal or compressive forcing are expected to appear mainly in regions of low Mach numbers (see Sect.~\ref{sec:mach}). The zooms were thus performed into the NGC\,2023 and NGC\,2024 star-forming regions (Fig.~\ref{fig:zoom})  where the Mach number lies between about 3 and 5 (see Fig.~\ref{fig:mach:ima}), mostly because the speed of sound is higher in regions of higher gas temperature, but also because the velocity dispersion is a bit lower.

Moreover, these regions offer the advantage of presenting a strongly localized emission. One of the requirements of the Brunt \& Federrath method is to use isolated fields. This is clearly not the case any more when zooming into these specific regions, and apodization is more necessary than ever to ensure that the signal falls smoothly to zero on the edges of the field. However, by using fields for which most of the signal is concentrated near the center, the effects of apodization are minimized, allowing us to stay as close as possible to the requirement of an isolated field. The smaller the field, the lower the value of $R$: this indicates an increasing proportion of compressive forcing.

\FigZoom{}

\section{Discussion}
\subsection{Compliance with the method's assumptions}
\label{sec:compliance}
In order to apply the Brunt \& Federrath method to real observational data, we were able to overcome several difficulties and sources of uncertainty.

First, the compliance of the dataset with the requirements of the method must be checked, namely the isotropy of the studied cloud, and its isolation. The whole field and the zooms present two opposite situations. The isolation criterion is well met in the case of the whole field: we have almost no signal to the West and to the South of the field, and very diffuse regions to the North and to the East (see Fig.~\ref{fig:moments}, first column), so that there is almost no need for apodization in order to have the signal going down to zero on the edges of the field. For the zooms, on the other hand, we are well into the cloud, so that there is signal all the way to the edges of the field. However, since we study local maxima of the emission, most of the signal is in the center, which minimizes the effects of the necessary apodization on the final results (only 7.9\% of lost signal for the deepest NGC\,2023 zoom, 7.6\% for NGC\,2024).

As far as the isotropy criterion is concerned, the 2D projection is quite obviously isotropic in the case of the zooms, since the considered fields are square, and much less in the case of the whole field, in which the region with signal has an aspect ratio of about 2:1. The third dimension is unknown, and in any case cannot match simultaneously the dimensions of the whole field and those of the deepest zooms. However, for the large diffuse regions like for the bright, compact regions, the dimension along the line of sight is supposed to be of the order of the dimensions in the plane of the sky. In the case of a zoom on a bright region embedded in a diffuse one, the signal, and therefore the dimension on which it is emitted along the line of sight, is dominated by the bright gas. Therefore, since the zooms are centred on a bright region, the corresponding datacube behaves almost as if the bright region was isolated and isotropic. Besides, the non-angle-averaged power spectra of $W_0$ and $W_1$ do not show any apparent anisotropy at any scale (except for the windowing  effects). If the power spectra are isotropic in two dimensions, then statistically we can expect the third dimension to follow this isotropy as well. Therefore, the isotropy requirement seems to be fulfilled by our dataset and the method can be applied.

Second, as was mentioned by \citet{brunt14}, the method is sensitive to values of the power spectra at large spatial scales (low frequencies of the power spectra) due to the characteristics of the sums in the parameters $A$ and $B$. We therefore had to find a way to obtain a smooth and reliable function that would represent an angle-averaged power spectrum at all frequencies. Once the power spectra were binned and fitted, we have two versions of the spectra, each with its flaws. The binned (``data only'') spectrum suffers from observational effects (noise and beam) but has also larger uncertainties at low spatial frequencies. The fitted spectrum, if extrapolated to all spatial frequencies (``fit only''), can give unphysical results in the lowest frequencies because they lie outside the power law validity range. For example, if the field had been zero-padded all around to create a very large square, such an extrapolation would give very high values of the spectrum at low frequencies, whereas physically they should be very low, since the field would on average be almost empty. These flaws led us to choose the composite scheme described in Sect.~\ref{sec:powerspectra}, where the fit result is used only in the ``inertial domain'', and the low frequencies keep the angle-averaged power spectrum as it is. Using a different version of the power spectra leads to quite different final results, as illustrated in Fig.~\ref{fig:compare}. However, it is important to note that even though the absolute value of the solenoidal fraction $R$ varies, the relative variations are consistent across scales whatever the power spectra computing scheme. Thus, results such as the unusually high solenoidal fraction at the scale of the whole cloud, or the variations of $R$ when zooming out of the star-forming regions, stay valid, and are further discussed below.
\FigCompare{}

\subsection{Physical interpretation}
To our knowledge, this work is the first attempt at applying the Brunt \& Federrath method to actual observational data \citep{brunt14, lomax15}. The results need to be compared with what was done so far on numerical simulations.

Both \citet{schneider13} and our calculations (see Sect.~\ref{sec:mach} and Appendix) show that we are in a context of highly supersonic turbulence, with a mean Mach number of about 6. We therefore expect a full mixing of the turbulence modes, so that the energy equipartition would predict a solenoidal fraction of $R = 2/3$ and a compressive fraction of $1 - R = 1/3$. Deviations from this ratio can either be a sign of a specific forcing for the turbulence in the case of transonic Mach numbers, as shown in the case of simulations \citep{brunt14}, or indicate that an ordered flow is superimposed on top of the turbulent flow.

The global value of $R > 0.72 \pm 0.09$ and the values on Fig.~\ref{fig:slide} can agree, within the error bars, with the expected value $R = 2/3$ in case of equipartition. However, the value $R > 0.72 \pm 0.09$ is still quite high. It can be the sign of a deviation in favour of solenoidal modes. At high Mach numbers, it would imply that an ordered solenoidal flow is superimposed on top of the turbulence. And Figure \ref{fig:moments} (column 2) shows that there is a large-scale differential motion in the cloud with the southern part receding while the northern part is approaching. This velocity shift could be the sign of a large-scale rotation of the whole cloud, which could dominate the smaller-scale motions and be responsible for the large fraction of solenoidal modes.

The fact that the turbulence in the Orion\,B molecular cloud is, on large scales, mostly solenoidal, is in agreement with the fact that, for its mass, Orion\,B has a low star-formation rate \citep{lada10}. Estimations of its SFE vary, with values ranging from 0.4\% to 3\% \citep{lada92, carpenter00, federrath13, megeath16}, but all studies show that the SFE in Orion\,B is about four times lower than in the neighbouring cloud Orion~A. In general, Orion\,B's SFE is regarded as particularly low, with \citet{megeath16} stressing that is has the lowest SFE among all molecular clouds closer than 500\pc. This remarkable feature of the Orion\,B cloud could be partially explained by the solenoidal flows that drive its velocity field, and hinder collapsing motions that could trigger star formation.
\\

In contrast, Figure \ref{fig:zoom} shows a major deviation from the equipartition in favour of compressive motions in two specific regions. When zooming deeply into the star-forming regions NGC\,2023 and NGC\,2024, we obtain solenoidal fractions as low as $R = 0.25$. The high fraction of compressive flow in these two regions most likely results from the infall of matter onto the star-forming region, and/or the expansion of the \Hii{} regions around the young massive stars \citep{tremblin14a, geen15}. The \Hii{} regions themselves are not observed in molecular tracers, so that this expansion is detected indirectly through the compression of the molecular gas at the ionization front of the \Hii{} regions.

For both regions, $R$ grows when zooming out, and will eventually reach the average values displayed on Fig.~\ref{fig:slide}. $R$ tends to decrease when reaching a field size of about $20'$. This behaviour can be due to the fact that the other star-forming region is entering the field of view, since the distance between the two cores is about $22'$, but it can also be related to the geometry of each region. 

In the case of NGC\,2023, this size of $20'$ also corresponds to the PDR of the Horsehead Nebula coming into the field, and, due to the pressure at the photo-dissociation front, it is expected to be a compressive region \citep{wardthompson06}, which is proven by the detection of at least one young star and one protostar in this region \citep{megeath12}. 

In the case of NGC\,2024, we see that $R$ starts increasing again after $25'$, even though NGC\,2023 comes more and more into the field. The variations of $R$ around $20'-25'$ might therefore be related to the location of the edge of the \Hii{} region around NGC\,2024. This \Hii{} bubble exerts a pressure on the surrounding gas \citep{tremblin14a,tremblin14b}, and this edge is therefore a highly compressive region, which can be seen in the form of an arc north of NGC\,2024 \citep{megeath12}. This region has a far lower surface density of stars that the inner part of NGC\,2024, but it is likely to be younger that NGC\,2024 (since it is a consequence of the expansion  of the \Hii{} region), and therefore might have formed only very young protostars, poorly detected by \emph{Spitzer} \citep{megeath16}, or no stars yet -- but it might become a very active region in the future.
\\

The sharp contrast between the large-scale solenoidal flow and the highly compressive flows in the NGC\,2023 and NGC\,2024 nebulae is in agreement with the spatial distribution of star formation observed in Orion\,B: \citet{lada92} observes that the large-scale SFE in the Orion\,B cloud is an order of magnitude lower than in the most massive cores, \citet{carpenter00} shows that, at his detection level, 100\% of young stars in Orion\,B are located in clusters (NGC\,2068 in the North, and NGC\,2024 in the South), which is unusual among the studied clouds, and \citet{lada13} conclude that Orion\,B is very ineffective at forming stars at $\Ak < 2.0$ mag, as compared to other GMCs. 

In a broader perspective, not only do these variations of $R$ confirm observationally the intuitive link between compressive motions and star formation, as proposed in simulations \citep{federrath12,padoan14}, but they also show that there can be an intra-cloud variability of the solenoidal fraction, in addition to the inter-cloud one. This shows that the large-scale environment of the cloud, although it plays a major role in driving the turbulence of the molecular cloud, cannot explain all by itself the repartition of solenoidal and compressive motions in the cloud: any denser region created by the density fluctuation in the compressible turbulent gas \citep[\eg{}][and references therein]{nolan15} can lead, under the effect of self-gravity or stellar feedback, to the formation of very localized, strongly compressive regions, even in the context of a mostly solenoidal flow. There is therefore no universal solenoidal fraction that can be applied generally to all clouds, and there are even intrinsic variations of $R$ from region to region within a cloud.

\section{Conclusion}
From a practical point of view, our work has shown that it is possible to apply the numerical method of \citet{brunt14} to observational data, in order to determine the fraction of solenoidal and compressive motions in a molecular cloud, using molecular lines as tracers of the density and velocity fields. We were able to pinpoint the observational requirements to apply this method to a dataset.

The spatial dynamic range is an important element, mostly in order to provide good quality power spectra. The field must be large enough to provide good statistics at low spatial frequencies, but it must also have a good spatial resolution, so that the power spectra have enough points to correct properly for the beam and noise effects. We found our minimum field size to be of at least 50 independent pixels.

In addition to the spatial resolution, having many independent spectral channels is of great help when correcting for the beam and noise effects. The spectral resolution must also be sufficient to resolve the studied spectral line.

The signal-to-noise ratio (SNR) also proved to be a key element during the calculations. An average SNR of at least 5 is desirable: our datacube, which has a mean SNR of 7.8, yielded a relative observational uncertainty of 13\% on the fraction of momentum in the solenoidal modes, and Figures \ref{fig:slide} and \ref{fig:zoom} show that this observational uncertainty contributes significantly to the overall uncertainty, and is even dominant at low solenoidal fractions.

The last point of the computation -- the density-velocity correlation, which also significantly contributes to the overall uncertainty, requires to use many spectral tracers of various typical densities. In that case, a spectral survey such as the one of the Orion\,B project is invaluable, in so far as all the needed tracers are available and inter-calibrated.
\\

From a physical point of view, the measurements have shown that the motions in the Orion\,B molecular cloud are highly supersonic, with a mean Mach number of $\sim 6$. However, the Mach number maps show large variations, with some regions being only moderately supersonic. These variations are due both to the variations of the temperature and to the turbulent velocity distribution in the cloud.

The largest scales of the cloud seem to be dominated by a rotational motion, which can be identified by a high solenoidal fraction in the flow. At smaller scales, we have shown that the motion is largely dominated  by compressive (infall and / or outflow) motions in the vicinity of the NGC\,2023 and NGC\,2024 star-forming regions. The northern edge of NGC\,2024 and the photo-dissociation front of the Horsehead nebula are also likely to be highly compressive regions, according to our results. This is in agreement with the observations of the star-formation efficiency in Orion\,B, which is unusually low at the scale of the whole cloud, and exclusively concentrated in clusters (NGC\,2023 and NGC\,2024).

The example of Orion\,B also shows that the star formation efficiency in a molecular cloud does not only depend on its overall fraction of momentum in the solenoidal modes of turbulence, but also on the local variations of this fraction, which can be driven by internal phenomena such as self-gravity and stellar feedback.
\\

This method could be applied in the future to study the variations of the solenoidal fraction between different molecular clouds, or between different regions or different chemical tracers within a given cloud. In the case of Orion\,B, we intend to analyse other data cubes for tracers such as C$^{18}$O or HCO$^+$, which trace respectively more compact and more diffuse regions of the cloud, in order to probe different environments in terms of density and temperature.

\begin{acknowledgements}
We wish to thank Patrick Hennebelle, Edith Falgarone and Pierre Lesaffre for fruitful conversations on hydrodynamics and magnetohydrodynamics, as well as our referee for enlightening remarks on compressible turbulence in the interstellar medium. This research also has made use of data from the Herschel Gould Belt survey (HGBS) project (http://gouldbelt-herschel.cea.fr). The HGBS is a Herschel Key Programme jointly carried out by SPIRE Specialist Astronomy Group 3 (SAG 3), scientists of several institutes in the PACS Consortium (CEA Saclay, INAF-IFSI Rome and INAF-Arcetri, KU Leuven, MPIA Heidelberg), and scientists of the Herschel Science Center (HSC). This work was supported by the CNRS/CNES program ``Physique et Chimie du Milieu Interstellaire'' (PCMI). We thank the CIAS for its hospitality during the two workshops devoted to this project. VVG thanks for support from the Chilean Government through the Becas Chile program. PG's postdoctoral position was funded by the INSU/CNRS. PG thanks ERC starting grant (3DICE, grant agreement 336474) for funding during this work. NRAO is operated by Associated 
Universities Inc. under contract with the National Science Foundation.

\end{acknowledgements}

\begin{appendix}

\section{Mach number estimation}
In this Appendix, we describe the computations that yielded the Mach number maps that allow us to estimate if the studied regions of the molecular clouds were below or above the hypersonic limit which is characterized by the equipartition of energy in the solenoidal and compressive modes of turbulence. We first compute the speed of sound using two different temperature maps, then derive the speed of the flow from the $^{13}$CO($J=1-0$) position-position-velocity cube, and finally obtain statistics on the 3-dimensional Mach number.

\subsection{Speed of sound}
The speed of sound in the gas needs to be derived from a temperature map, be means of $c_\emr{sound} = \sqrt{\gamma\cdot R_\emr{gas}\cdot T/m_\emr{mol}}$, where $\gamma$ is the adiabatic index, $R_\emr{gas}$ the universal gas constant and $m_\emr{mol}$ the molar mass of the gas.

We have directly access to two temperature maps: the dust temperature, computed by the \emph{Herschel} Gould Belt Survey consortium \citep{schneider13}, and the peak temperature of the $^{12}$CO map, from our IRAM-30m observations. The dust temperature is expected to be a lower bound for the kinetic temperature, because dust radiates more efficiently than gas. The gas and dust become coupled only for densities larger than $10^4$\pccm{}  \citep{goldsmith01}.

For optically thick gas, the excitation temperature $T_\emr{exc}$ is determined \citep[\eg{} in][]{rohlfs04} by
\begin{equation}
\exp\left(\frac{h \nu}{k_\emr{B} T_\emr{exc}}\right) -1 = \left(\frac{T_\emr{peak}}{h\nu/k_\emr{B}} + \frac{1}{\exp\left(\frac{h \nu /k_\emr{B}}{T_\emr{CMB}}\right)-1}\right)^{-1}
\end{equation}
where $\nu = 115.271202\GHz$ is the frequency of the $^{12}$CO($J=1-0$) transition, and $T_\emr{CMB} = 2.728$ K is the cosmic microwave background temperature. We assume that the excitation temperature is close to the kinetic temperature of the gas (local thermodynamic equilibrium).

We assume that at very low column density (and therefore low $T_\emr{peak}$), the kinetic temperature is underestimated (since the computation is only valid for optically thick gas). The dust temperature is supposed to be a lower limit close to the kinetic temperature of the gas, except when it is not deemed reliable any more: above a threshold of 60\,K, we deem that \emph{Herschel} and \emph{Planck} observations do not yield the best temperature estimate due to their limited wavelength ranges ($50-600$ \micron{} for \emph{Herschel}). At these temperatures, $^{12}$CO($J=1-0$) is usually saturated enough to give a good result for the excitation temperature, as can be seen from the $^{12}$CO($J=1-0$)/$^{13}$CO($J=1-0$) line ratio which diverges significantly from the $^{12}$C/$^{13}$C abundance ratio \citep{pety16,ripple13}.  We can therefore construct a third temperature map $T_\emr{max}$ using the dust temperature and the excitation temperature of $^{12}$CO($J=1-0$): we use the excitation temperature whenever it is above the 60\,K threshold, and in other regions, we use the maximum of the gas excitation temperature and the dust temperature (Fig.~\ref{fig:soundspeed}, panels 1 to 3).

\FigSoundSpeed{} %

When computing the speed of sound from the temperature map, the nature of the gas comes into play. We considered a 75\% -- 25\% mixture in mass of molecular hydrogen and helium, which yields a molecular mass of 2.513\,kg.mol$^{-1}$. Since we are not simply dealing with a mono-atomic or diatomic ideal gas, the adiabatic index $\gamma$ of this mixture has to be determined. To compute $\gamma$, we used tabulated NIST values of the calorific capacities $C_P$ and $C_V$ of the two gases at an average temperature of 24 K. The resulting value, $\gamma = 1.66674$, is very close to the value 5/3 that would be expected for a mono-atomic ideal gas -- which could be expected since at such low temperatures, the ro-vibrational modes of H$_2$ are frozen.

\subsection{Flow velocity}
The turbulent velocity dispersion was computed using the velocity dispersion along each line of sight, which can be derived from the $W_2$ map. We determined the flow velocity dispersion as $u = \sqrt{W_2/W_0}$ (Fig.~\ref{fig:soundspeed}, last panel).

This computation implies two assumptions. First, we assume that the natural and thermal widths of the lines are negligible, compared to the turbulent broadening. Second, we neglect the opacity broadening as well. While it is correct that thermal broadening has less that 1 \% effect on the line width, the assumption for the opacity broadening is less obvious. We determined that the expected correction would be of the order of 10 \% for the brightest lines of sight, reducing the width of the lines and therefore the Mach number. However, given that the correction would have been difficult to implement for non-Gaussian lines, as it is the case in our $^{13}$CO($J=1-0$) field, and that the correction would be significant only for a small fraction of the lines of sight, we left the turbulent velocity field uncorrected.

\subsection{Results}
The ratio of the turbulent velocity to the sound speed gives us the Mach number along the $z$ axis. The total Mach number is $M = \sqrt{3}M_z$, when the turbulence is isotropic. The Mach number is determined twice, using two maps: the excitation (gas) temperature, and the maximum of dust and excitation temperatures as described above. We then compare the results of the two computations (Fig.~\ref{fig:mach:ima} and \ref{fig:mach:hist}).

\FigMachImage{}%

\FigMachHist{}%

To estimate the typical Mach number at the scale of the whole cloud, we plot the histogram of the obtained maps (Fig.~\ref{fig:mach:hist}). The shapes of the histograms are very similar for $T_\emr{max}$ and $T_\emr{exc}$. Due to the shape of both distributions, with a large tail at high Mach number, the most probable value of the Mach number for both distributions ($M = 3.5$), is significantly smaller than the mean and the median values, as shown in Tab.~\ref{tab:mach}.

\TabMach{}

\end{appendix}

\bibliographystyle{aa} %
\bibliography{turbulence} %

\end{document}